\documentclass[fleqn,usenatbib,useAMS]{mnras}

\usepackage{graphicx}	
\usepackage{amsmath}	
\usepackage{amssymb}	
\usepackage{multicol}   
\usepackage{bm}		    
\usepackage{pdflscape}	
\usepackage{epstopdf}
\usepackage{longtable}
\usepackage{booktabs}
\usepackage{newtxmath}
\usepackage[a4paper]{geometry}
\usepackage{multirow}
\usepackage{pdflscape}
\usepackage{tabularx}
\usepackage{layouts}
\newcolumntype{Y}{>{\centering\arraybackslash}X}

\newcommand{\pc}{{\rm\thinspace pc}}
\newcommand{\cm}{{\rm ~ cm}}
\newcommand{\km}{{\rm ~ km}}

\newcommand{\metre}{{\rm ~ m}}
\newcommand{\kmps}{\hbox{${ ~ \rm\km\s^{-1}}$}}
\newcommand{\cmc}{\hbox{${ ~ \rm\cm^{-3}}$}}
\newcommand{\kpc}{{\rm ~ kpc}}

\newcommand{\s}{{\rm ~ s}}
\newcommand{\au}{{\rm ~ au}}

\newcommand{\yr}{{\rm ~ yr}}

\newcommand{\beam}{{\rm ~ beam}}
\newcommand{\mJy}{{\rm ~ mJy}}
\newcommand{\uJy}{{\rm ~ \upmu Jy}}

\newcommand{\Msol}{\hbox{${\rm ~ M_{\odot}}$}}
\newcommand{\Lsol}{\hbox{${\rm ~ L_{\odot}}$}}
\newcommand{\Lbol}{\hbox{${\rm ~ L_{Bol.}}$}}
\newcommand{\HII}{\text{H\,\textsc{ii}}}

\newcommand{\rahr}{\hbox{$\rm^h$}}
\newcommand{\ramin}{\hbox{$\rm^m$}}
\newcommand{\rasec}{\hbox{$\rm^s$}}
\newcommand{\GHz}{{\rm ~ GHz}}
\newcommand{\MHz}{{\rm ~ MHz}}

\title[Temporal studies of massive ionized jets]{Investigating the temporal domain of massive ionized jets - I. A pilot study}

\author[S. J. D. Purser, S. L. Lumsden, M. G. Hoare and N. Cunningham]{S. J. D. Purser$^{1}$\thanks{E-mail:
pysjp@leeds.ac.uk}, S. L. Lumsden$^{1}$, M. G. Hoare$^{1}$ and N. Cunningham$^{2}$\\
$^{1}$School of Physics and Astronomy, University of Leeds, Leeds LS2 9JT, UK\\
$^{2}$National Radio Astronomy Observatory, PO Box 2, Green Bank, WV 24944, USA}
\date{}
\pubyear{2017}

\begin{document}
\label{firstpage}
\pagerange{\pageref{firstpage}--\pageref{lastpage}}
\maketitle

\begin{abstract}
We present sensitive ($\sigma<10\uJy\beam^{-1}$), radio continuum observations using the Australian telescope compact array (ATCA) at frequencies of $6$ and $9$ GHz towards 4 MYSOs. From a previous, less sensitive work, these objects are known to harbour ionized jets associated with radio lobes, which result from shock processes. In comparison with that work, further emission components are detected towards each MYSO. These include extended, direct, thermal emission from the ionized jet's stream, new radio lobes indicative of shocks close ($<10^5\au$) to the MYSO, 3 radio Herbig-Haro objects separated by up to $3.8\pc$ from the jet's launching site and an IR-dark source coincident with CH$_3$OH maser emission. No significant, integrated flux variability is detected towards any jets or shocked lobes, and only one proper motion is observed ($1806\pm596\kmps$ parallel to the jet axis of G310.1420+00.7583A). Evidence for precession is detected in all four MYSOs with precession-periods and angles ranging from $66-15480\yr$ and $6-36\degr$ respectively. Should precession be the result of the influence from a binary companion, we infer orbital radii of $30-1800\au$. 

\end{abstract}


\section{Introduction}
\label{sec:intro}

Considering the importance of massive ($>8\Msol$) stars for astrophysics in general, little is known for certain about the processes involved in their formation. The low mass formation paradigm, in contrast, is relatively well understood \citep{shu1987a}. Fundamentally, inside-out collapse of a core results in the formation of a central protostar which, due to angular momentum considerations, forms a flattened disc of infalling material. Inward motion in the disc results in accretion of this material onto the forming star, with a typical rate (in the low-mass case of a $1\Msol$ accreting YSO) of $\sim10^{-7}\Msol\yr^{-1}$ \citep{Gullbring1998}. Typical rates in high mass stars, however, need to be much higher $\sim10^{-4}\Msol\yr^{-1}$ to accommodate for short pre-main sequence phases \citep[$\sim10^5\yr$,][]{Mottram2011b}. Thus, debate over how closely the accretion process mimics that of the low mass case still ensues. 

One ubiquitous feature of massive star formation are the massive molecular outflows \citep{Beuther2002} which are detected in association with their powering massive young stellar objects (MYSOs). These outflows are thought to be mechanically coupled to jets of material, ejected perpendicular to the disc during accretion, which seems to be a common phenomenon in both low mass \citep{Anglada1996} and high mass protostars \citep[incidence rates of $\sim65\%$,][from now on referred to as Paper I]{Purser2016}\defcitealias{Purser2016}{Paper~I}. Furthermore, the ejection of jets is thought to be linked to inflow processes, with fixed accretion-to-ejection ratios for some models of jet launching/collimation \citep[see section 2.1 in the review by][and references therein]{Frank2014PPVI}. Therefore the exact nature and morphology of the jets themselves can indirectly inform us on the accretion process.

From previous observations of ionized jets, their general, radio morphology seems to take the form of an elongated component with a thermal, spectral index centred on the MYSO. Theoretical studies indicate that the value for the spectral index can range from $-0.1$ to $1.4$ \citep{Reynolds1986}, with observational studies seemingly averaging a spectral index of $\sim0.6$ \citep[with ranges from $0.2$ to $0.9$ in the cases of HH 80-81 and G345.4938+01.4677 respectively;][]{Marti1993,Guzman2010}. In $\sim50\%$ of cases this is associated with separate lobes of emission \citepalias{Purser2016}, which are presumably a result of the jet ionizing and shocking the ambient material of the envelope/clump, or internal shocks within the jet as a result of variability in ejection velocity \citep[such as in the low mass case of HH 211,][]{Moraghan2016}. These lobes tend to be spatially distinct and separate in the majority of cases \citepalias[average separations of $\sim10^4\au$,][]{Purser2016}, with little radio emission seen in between thermal and non-thermal components e.g. IRAS 16547-4247 \citep{Rodriguez2005} and IRAS 16562-3959 \citep{Guzman2010}.  This suggests one of two possibilities, that the ejection of material is highly variable/episodic \citep[possibly a result of fragmentation in the accretion disc,][]{Meyer2017}, or that a constant, collimated outflow of material only sporadically impinges upon the surrounding matter to form shocks.

In cases where multiple lobes are seen, it is common to find a positional offset for the lobe from the overall position angle of the jet, suggesting the outflow axis evolves over time. Precession of collimated outflows are predicted in the simulations of \citet{SheikhnazamiFendt2015} who performed 3D MHD simulations of a disc-jet system influenced by the gravitational potential of a binary companion. They find that the tidal interactions for separations of $\sim200$ inner disc radii ($\sim2000\au$ in reality) warp the accretion disc of the YSO, resulting in disc, and therefore jet, precession. Considering the high companion fraction observed towards massive stars \citep[see section 3.5 of][and references therein]{Duchene2013} we might expect jet precession to be a relatively common phenomenon. Interestingly their simulations also predict variable accretion and outflow rates which would be seen in observations as flux variability \citep{AMI2011a}, an effect recently observed towards a $20\Msol$ MYSO \citep{Caratti2017}.

Radio observations of both variability and precession of massive jets are present in the literature, but are few and far in between. For example, \citet{Rodriguez2008} observed the MYSO IRAS 16547-4247 at two epochs separated by $\sim1000$ days. From comparison of the flux maps in both epochs, no proper motions along the outflow axis were seen, however precession was apparent in the non-thermal lobes with a rate of $0.08\,\degr\yr^{-1}$ and therefore a derived period of $<4500\yr$. Both the central jet and a non-thermal lobe (component S-1) increased in flux over time by $9\%$ and $36\%$ respectively. As far as proper motion observations, several works have reported radio-lobe velocities ranging from $300-1000\,\mathrm{km\,s}^{-1}$ in massive cases \citep[such as HH80-81 and Cep A HW2 from][respectively]{Marti1998,Curiel2006}.

Considering the discussion above, this work aims to investigate how ionized jets change over time. Specifically we examine their flux variability, precession and proper motions. To do this we build upon the results of a previous survey of ionized jets \citepalias{Purser2016} which was based upon a well-selected sample of MYSOs from the RMS (Red MSX Source) survey \citep{Lumsden2013}. In that work 26 ionized jets were detected from a sample of 49 objects, of which 11 of the 26 were associated with non-thermal emission. Using a new set of radio observations towards 4 of these 11 sources, this work increases the sensitivity by a factor of $2-3$ ($\sim10\uJy\,\mathrm{beam}^{-1}$) in comparison to \citetalias{Purser2016}. Specific questions we aim to answer are, do the radio jets exhibit variability over the period of time between observations ($\sim2\yr$)? Despite these short time-baselines, can we still detect large proper motions towards the jets? Do they show precession in their propagation axes? Is there any fainter emission previously undetected? In light of answers to the previous questions, can we further constrain models of massive star/jet formation? 

\section{Sample and Observations}
\label{sec:obs}

As mentioned above, 4 MYSOs associated with ionized jets were chosen from the sample of \citetalias{Purser2016} to be re-observed. These objects were selected on the basis of the presence of shock-ionized lobes, for which proper motions could be investigated, as well as a thermal jet centred on the MYSO. A range in bolometric luminosities are present across the four objects, and the lowest luminosity object (G263.7434+00.1161) was included on the basis of its hour angle, which allowed it to be observed at times when the rest of the sample were inaccessible. Further to this, its bolometric luminosity is typical of intermediate-mass objects and therefore its inclusion in the sample also allows for the investigation of this transition region from low-mass to high-mass regimes.

\begin{table*}
\centering
\caption{A table of the positions, calibration types and fluxes for the calibrators used in the reduction of the data. The science target which the phase calibrators are used to transfer complex gain solutions to are also listed (in an abbreviated form).}
\begin{tabular}{lcccccc}
\hline
\textbf{Calibrator} & \textbf{R.A.} & \textbf{Dec.} & \textbf{Type} & \textbf{Freq.} & \boldmath$\mathrm{S}_\nu$ & \textbf{Science Target(s)} \\
 & (J2000) & (J2000) & & (GHz) & (Jy) & \\
\hline
\multirow{2}{*}{0826$-$373} & \multirow{2}{*}{$08\rahr28\ramin04.78\rasec$} & \multirow{2}{*}{$-37\degr31\arcmin06.3\arcsec$} & \multirow{2}{*}{Phase} & 6 & $1.55\pm0.09$ & \multirow{2}{*}{G263.7434} \\
 &  &  &  & 9 & $1.30\pm0.10$ &  \\
\multirow{2}{*}{1352$-$63} & \multirow{2}{*}{$13\rahr55\ramin46.63\rasec$} & \multirow{2}{*}{$-63\degr26\arcmin42.6\arcsec$} & \multirow{2}{*}{Phase} & 6 & $1.22\pm0.14$ & \multirow{2}{*}{G310.0135, G310.1420, G313.7654} \\
 &  &  &  & 9 & $1.08\pm0.29$ &  \\
\multirow{2}{*}{1934$-$638} & \multirow{2}{*}{$19\rahr39\ramin25.03\rasec$} & \multirow{2}{*}{$-63\degr42\arcmin45.6\arcsec$} & \multirow{2}{*}{Bandpass, Flux} & 6 & $4.51\pm0.35$ &  \\
 &  &  &  & 9 & $2.72\pm0.16$ &  \\
\hline
\end{tabular}
\label{tab:CalInfo}
\end{table*}

\begin{table*}
\centering
\caption{A table of the target sources, their positions, associated IRAS sources, distances, bolometric luminosities, ZAMS stellar masses \citep[from the models of][assuming a $30\%$ error in $\Lbol$]{Davies2011}, total integration times and theoretical image noise levels per beam (utilising a robustness of 0) at 6 and $9\GHz$.}
\begin{tabular}{lccccccccc}
\hline
\textbf{Object} & \textbf{R.A.} & \textbf{Dec.} & \textbf{IRAS} & \textbf{D} & \boldmath$\Lbol$ & \boldmath$\mathrm{M_\star}$ & \boldmath$\tau_\mathrm{int.}$ & \boldmath$\sigma_6$ & \boldmath$\sigma_9$ \\
 & (J2000) & (J2000) & & (kpc) & $(\Lsol)$ & $(\Msol)$ & (hrs) & ($\uJy$) & ($\uJy$)\\
\hline
G263.7434+00.1161 & $08\rahr48\ramin48.64\rasec$ & $-43\degr32\arcmin29.0\arcsec$ & 08470$-$4321 & 0.7 & $1.2\times10^{3}$ & $6.3^{+0.6}_{-0.2}$ & 7.33 & 7.3 & 8.4 \\
G310.0135+00.3892 & $13\rahr51\ramin37.85\rasec$ & $-61\degr39\arcmin07.5\arcsec$ & 13481$-$6124 & 3.2 & $6.7\times10^{4}$ & $24.3^{+3.1}_{-3.2}$ & 6.19 & 7.9 & 9.1 \\
G310.1420+00.7583A & $13\rahr51\ramin58.27\rasec$ & $-61\degr15\arcmin41.7\arcsec$ & 13484$-$6100 & 5.4 & $8.0\times10^{3}$ & $11.2^{+1.0}_{-1.2}$ & 2.99 & 11.4 & 13.1 \\
G313.7654$-$00.8620 & $14\rahr25\ramin01.53\rasec$ & $-61\degr44\arcmin57.6\arcsec$ & 14212$-$6131 & 7.8 & $6.1\times10^{4}$ & $23.4^{+2.7}_{-3.0}$ & 5.29 & 8.6 & 9.9 \\
\hline
\end{tabular}
\label{tab:TargetInfo}
\end{table*}

All radio observations were made using the Australian Telescope Compact Array (ATCA) in the 6A configuration, on the 19th, 20th and 21st December 2014. For reference, the observations of \citetalias{Purser2016} were conducted from the 25th to 28th of February 2013. A total of 4 individual objects were observed at 2 different frequency bands (centred on 6.0 and $9.0\GHz$). These frequencies were observed using a bandwidth of $2048\MHz$ (XX, YY, XY, and YX polarizations) split evenly either side of the  central frequencies, which was subsequently divided into $1\MHz$ channels. From this point on the observed frequencies are referred to as the $6$ and $9\GHz$ bands.

The range of scales the instrument was sensitive to were $1.8-18.5\arcsec$ and $1.2-12.3\arcsec$ for the $6$, and $9\GHz$ frequency bands respectively, corresponding to a minimum baseline length of $337\metre$ and a maximum of $5939\metre$. Differing scale sensitivities can manifest in the results by a decrease in the amount of flux recovered at higher frequencies (i.e.\ on larger spatial scales). \citetalias{Purser2016} conducted synthetic observations towards an idealised jet model, representative of their sample in both spatial scale, and spectral index. That investigation showed that for the ATCA in the 6A configuration, $>92\%$ of the flux was recovered at $6$ and $9\GHz$, and the recovered spectral index did not vary from the idealised model. Thus, these effects can be neglected in the further analysis of this paper. Furthermore, the synthetic observations showed that the signal to noise ratio was insufficient to accurately recover the physical dimensions of the object. However, with the deeper integration times of this dataset, more accurate dimensions should be deconvolved from the data.

Scan times on the flux calibrator, phase calibrators and science targets were 8, 2 and 15 minutes respectively. In order to provide more coherent phase solutions between the two epochs, therefore increasing the reliability of image comparison and analysis, the phase calibrators were the same quasars as those used in \citetalias{Purser2016}. For the flux calibrator, an absolute flux scale uncertainty of $5\%$ is adopted for both frequencies. Listed in Tables \ref{tab:CalInfo} and \ref{tab:TargetInfo} are the observed calibrators and science targets. 

For reducing the data, the Multichannel Image Reconstruction Image Analysis and Display (MIRIAD) software package \citep{miriad} was used. 

In the event that methanol masers or bright ($>10\mJy$) continuum sources were present in the field of view, phase-only (due to ATCA's limited number of baselines) self-calibration was iteratively performed until no further improvement in the RMS scatter of the phase solutions was achieved. 

\begin{figure*}
\includegraphics[]{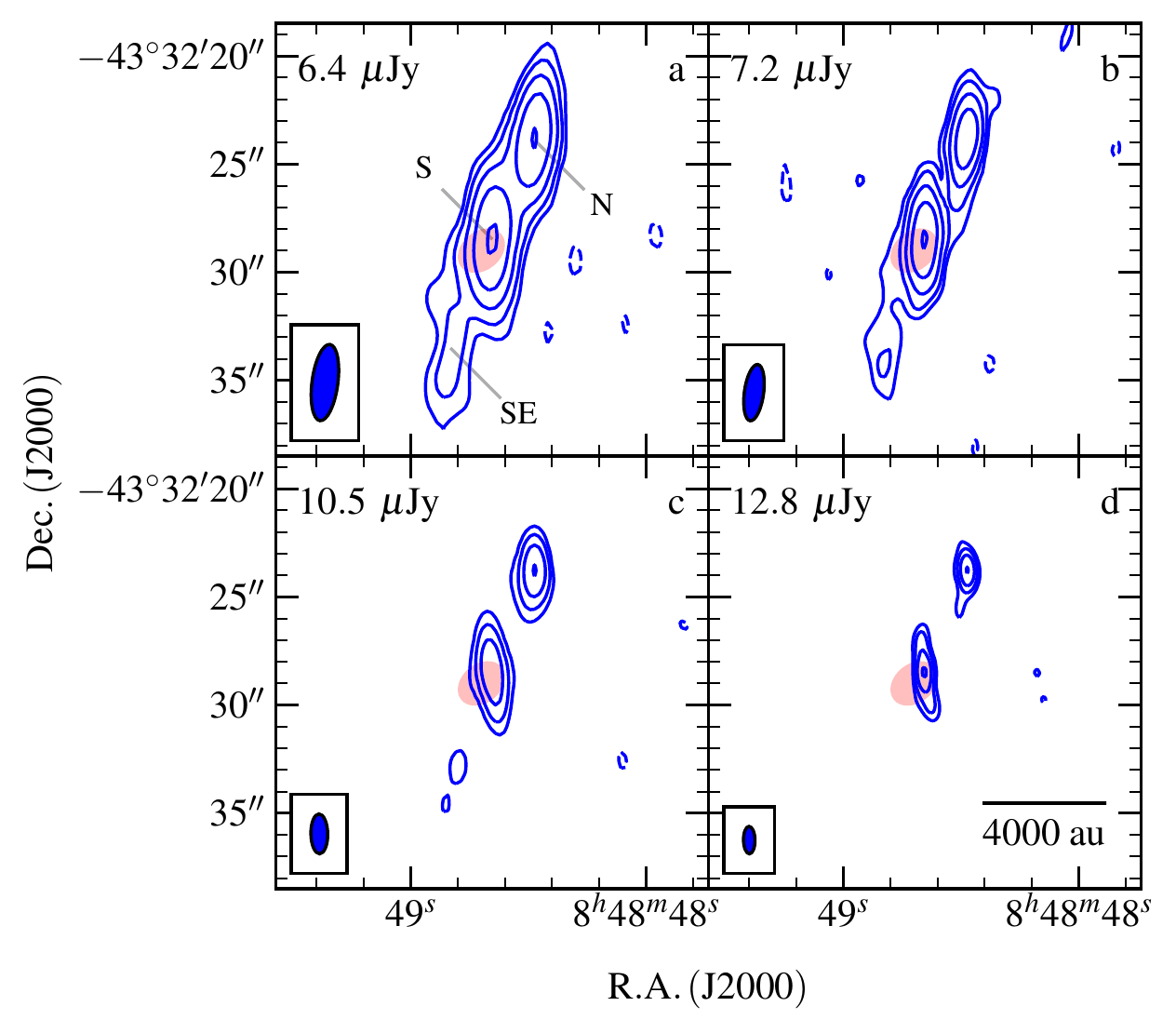} 
\caption{Contour plots of radio flux density for observations made towards G263.7434+00.1161. RMS noise is marked in the top-left of each of the sub-plots which are: a.) Image of 6 GHz data using a robustness of 0.5 where contours are $(-3, 3, 7, 15, 35, 79)\times \sigma$; b.) Image of 9 GHz data using a robustness of 0.5 where contours are $(-3, 3, 7, 15, 35, 78) \times \sigma$; c.) Image of 6 GHz data using a robustness of -1 with contours located at $(-3, 3, 7, 18, 42) \times \sigma$; d.) Image of 9 GHz data using a robustness of -1 with contours located at $(-3, 3, 7, 15, 33) \times \sigma$. Restoring beams are indicated in the bottom left corner of each plot and are $3.55\arcsec\times1.20\arcsec$ at $\theta_{PA}=-7.48\degr$, $2.62\arcsec \times 0.89\arcsec$ at $\theta_{PA}=-7.76\degr$, $1.84\arcsec \times 0.78\arcsec$ at $\theta_{PA}=  1.07\degr$ and $1.27\arcsec \times 0.53\arcsec$ at $\theta_{PA}=0.98\degr$ for sub-plots a, b, c and d respectively. The $3\sigma$ positional error ellipse for the MSX point source associated with the MYSO is plotted in red.}
\label{fig:g263_7434contour}
\end{figure*}

\begin{figure*}
\includegraphics[]{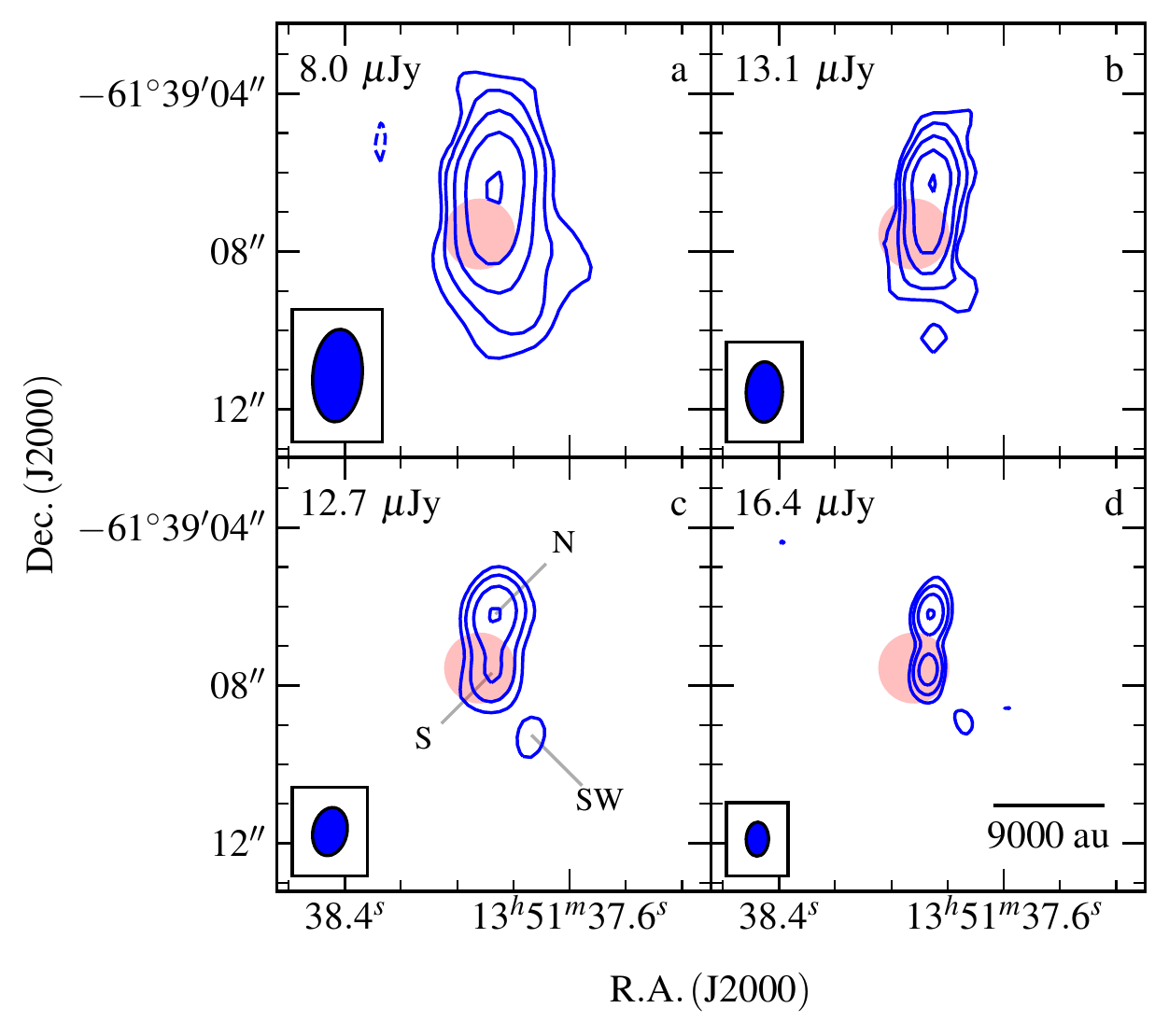} 
\caption{Contour plots of radio flux density for observations made towards G310.0135+00.3892. RMS noise is marked in the top-left of each of the sub-plots which are: a.) Image of 6 GHz data using a robustness of 0.5 where contours are $(-4, 4, 9, 20, 44, 97)\times \sigma$; b.) Image of 9 GHz data using a robustness of 0.5 where contours are $(-4, 4, 8, 14, 27, 52) \times \sigma$; c.) Image of 6 GHz data using a robustness of -1 with contours located at $(-4, 4, 9, 21, 49) \times \sigma$; d.) Image of 9 GHz data using a robustness of -1 with contours located at $(-4, 4, 8, 17, 36) \times \sigma$. Restoring beams are indicated in the bottom left corner of each plot and are $2.35\arcsec\times1.25\arcsec$ at $\theta_{PA}=-4.43\degr$, $1.53\arcsec \times 0.92\arcsec$ at $\theta_{PA}=-1.13\degr$, $1.23\arcsec \times 0.87\arcsec$ at $\theta_{PA}=-12.29\degr$ and $0.87\arcsec \times 0.57\arcsec$ at $\theta_{PA}=-1.84\degr$ for a.), b.), c.) and d.) respectively. The $3\sigma$ positional error ellipse for the MSX point source associated with the MYSO is plotted in red.}
\label{fig:g310_0135contour}
\end{figure*}

\begin{figure}
\includegraphics[width=\columnwidth]{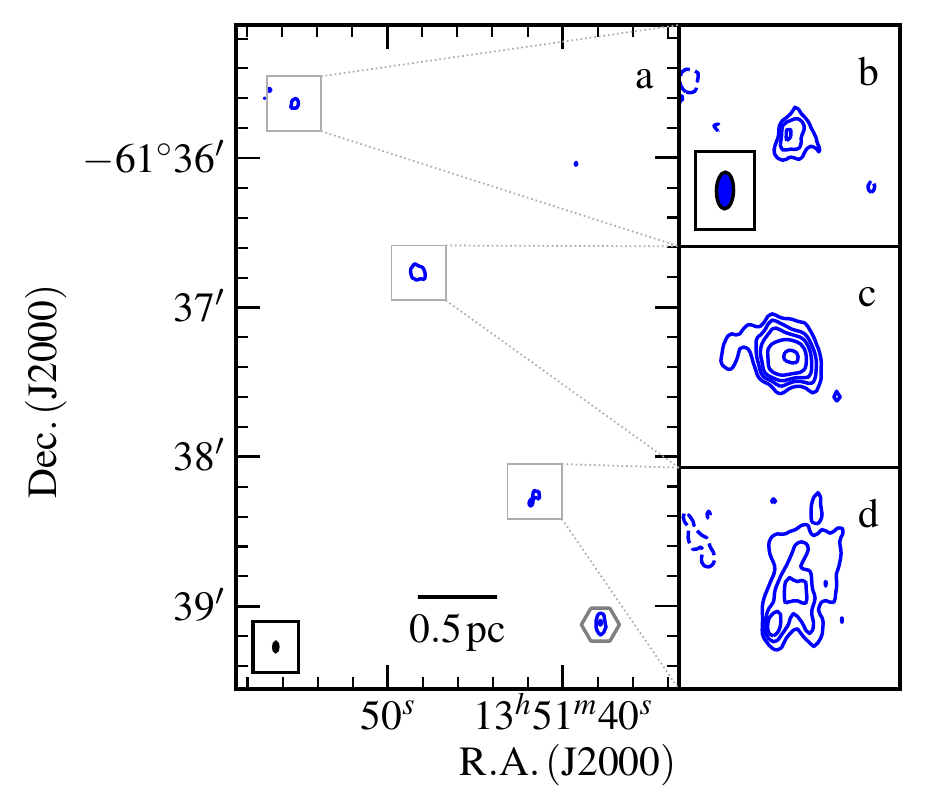}
\caption{Primary beam corrected $6\GHz$ image (robustness of $2$) of G310.0135+00.3892 and its associated radio Herbig-Haro objects. Each plot is: a.) the overall area presented with 4 and 50$\sigma$ contours where $\sigma=17.1\uJy\,\mathrm{beam}^{-1}$ with G310.0135+00.3892 is located in the bottom right (hexagonal marker); b.) Sub-plot of HH3 ($\sigma=17.1\uJy$); c.) Sub-plot of HH2 ($\sigma=13.3\uJy$); d.) Sub-plot of HH1 ($\sigma=10.4\uJy$). Contours are set at $(-3, 3, 5, 7, 10, 15)\times\sigma$ and the field of view is $22\arcsec$, for b, c and d.}
\label{fig:g310_0135_HH}
\end{figure}

\begin{figure*}
\includegraphics[]{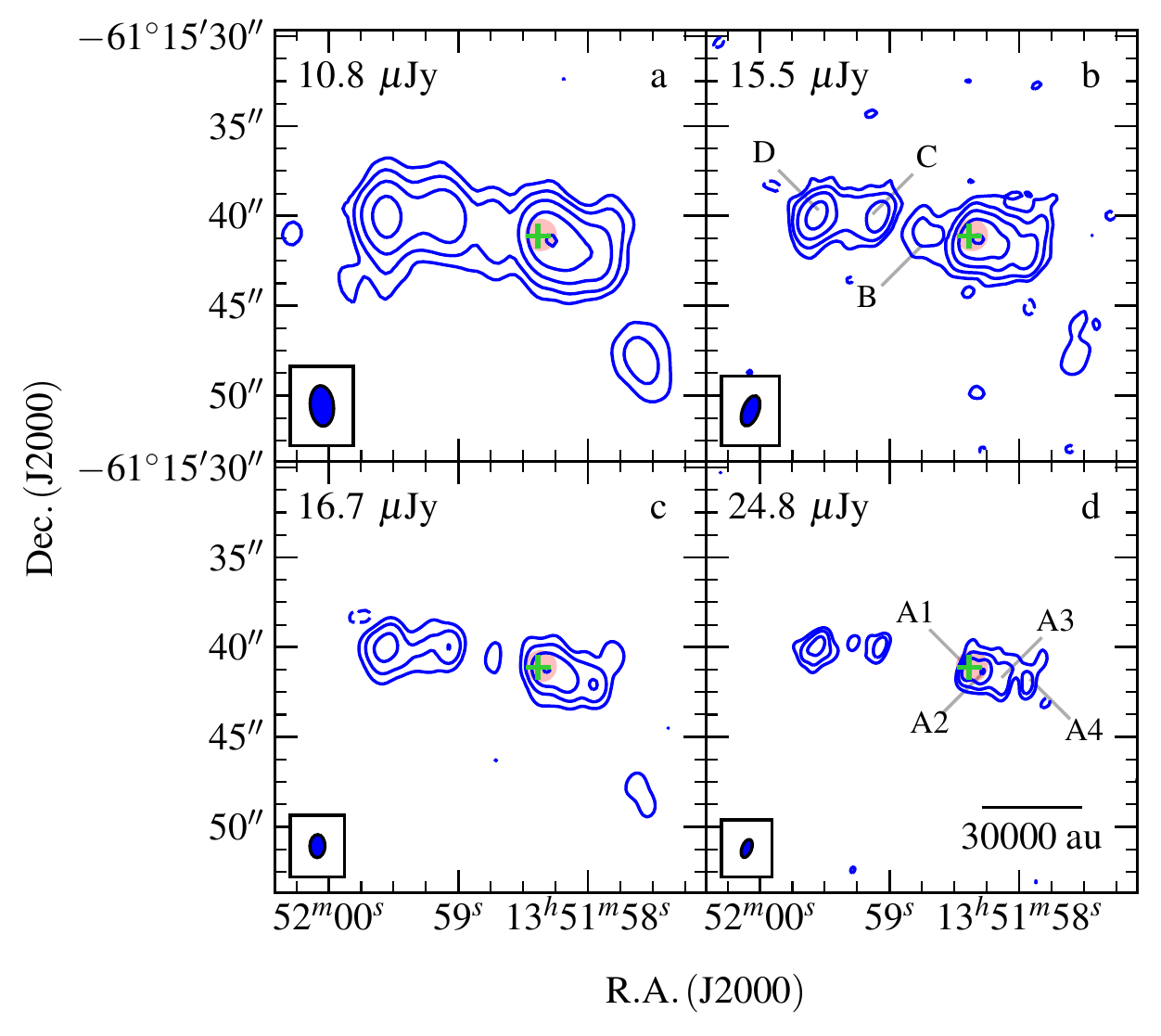}  
\caption{Contour plots of radio flux density for observations made towards G310.1420+00.7583A. RMS noise is marked in the top-left of each of the sub-plots which are: a.) Image of 6 GHz data using a robustness of 0.5 where contours are $(-3, 3, 6, 13, 27, 58)\times \sigma$. The restoring beam was $2.26\arcsec\times1.30\arcsec$ at $\theta_{PA}=5.73\degr$; b.) Image of 9 GHz data using a robustness of 0.5 where contours are $(-3, 3, 6, 13, 26, 54) \times \sigma$. The restoring beam was $1.77\arcsec \times 0.89\arcsec$ at $\theta_{PA}=-21.04\degr$; c.). Zoomed in image of 6 GHz data using a robustness of -1 with contours located at $(-3, 3, 7, 15, 32) \times \sigma$. The restoring beam was $1.28\arcsec \times 0.84\arcsec$ at $\theta_{PA}=-1.86\degr$; d.) Zoomed in image of 9 GHz data using a robustness of -1 with contours located at $(-3, 3, 6, 13, 27) \times \sigma$. The restoring beam was $1.05\arcsec \times 0.54\arcsec$ at $\theta_{PA}= -20.89\degr$. The $3\sigma$ error ellipse in the position of the MSX point source is shown in red, while the detected methanol maser is shown as a green cross.}
\label{fig:g310_1420contour}
\end{figure*}

\begin{figure*}
\includegraphics[]{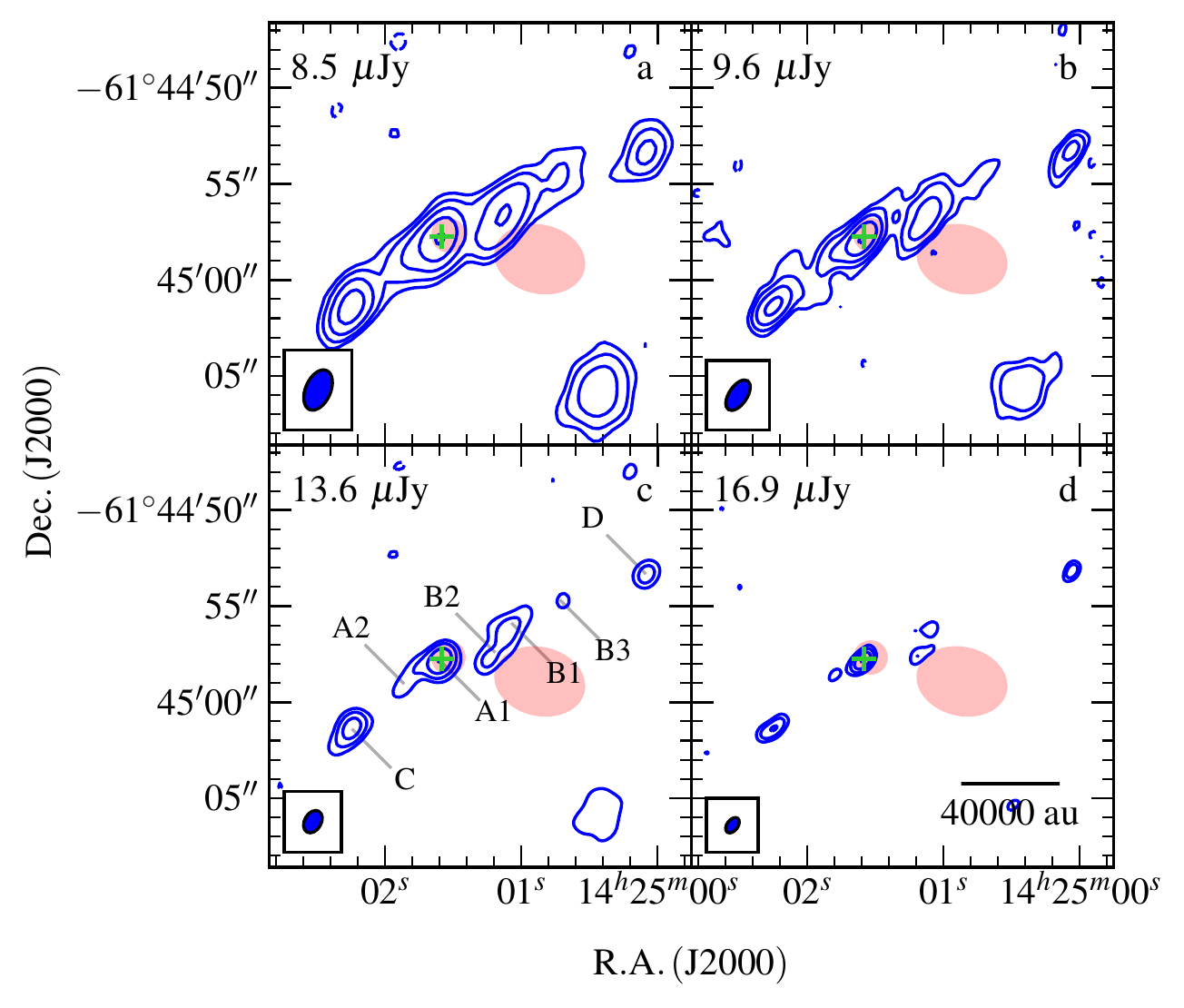} 
\caption{Contour plots of radio flux density for observations made towards G313.7654$-$00.8620. RMS noise is marked in the top-left of each of the sub-plots which are: a.) Image of 6 GHz data using a robustness of 0.5 where contours are $(-3, 3, 6, 13, 27, 58)\times \sigma$. A restoring beam of $2.22\arcsec\times1.31\arcsec$ at $\theta_{PA}=-23.69\degr$ was used; b.) Image of 9 GHz data using a robustness of 0.5 where contours are $(-3, 3, 6, 13, 26, 54) \times \sigma$. A restoring beam of $1.80\arcsec \times 0.95\arcsec$ at $\theta_{PA}=-33.80\degr$ was used; c.). Zoomed in image of 6 GHz data using a robustness of -1 with contours located at $(-3, 3, 7, 15, 32) \times \sigma$. A restoring beam of $1.25\arcsec \times 0.87\arcsec$ at $\theta_{PA}=-29.68\degr$ was used; d.) Zoomed in image of 9 GHz data using a robustness of -1 with contours located at $(-3, 3, 6, 13, 27) \times \sigma$. A restoring beam of $0.91\arcsec \times 0.57\arcsec$ at $\theta_{PA}= -37.09\degr$ was used. Both MSX (larger) and GLIMPSE error ellipses (smaller) are shown in red. A green cross marks the position of a detected methanol maser.}
\label{fig:g313_7654contour}
\end{figure*}

\section{Results}
\label{sec:results}
Maps of radio flux towards each of the 4 objects in our sample are shown in Figures \ref{fig:g263_7434contour} to \ref{fig:g313_7654contour} with separate clean maps produced at each frequency and each utilised robustness of $-1$ (improved resolution, reduced effects from residual side-lobes) and $0.5$ (increased flux/large-scale sensitivity). Any channels displaying strong maser emission in the $6\GHz$ band were imaged and the positions of the brightest maser spot plotted as a green cross in the relevant contour plots. Values for the integrated fluxes (derived using the \textsc{casa} task \textsc{imfit}), spectral indices ($\alpha$) and power-law coefficient for the variation of the deconvolved major-axis length with frequency ($\gamma$) are presented in \autoref{tab:JetsMainProperties} for each identified radio lobe. In each case, the robustness of the clean maps used to derive these quantities is listed, with spatially-distinct lobes generally utilising a natural robustness and more confused sources making use of the enhanced resolution of the more uniformly weighted clean map (at the cost of increased noise).

We now move on to give brief reviews of previous observations and discuss the results of the 2014 images, on an object-by-object basis.

\begin{table*}
\centering
\caption{A table of the integrated fluxes (whose errors take into account a $5\%$ absolute flux error at $6$ and $9\GHz$), derived spectral indices for flux $(\alpha)$ and major axis length $(\gamma)$, for all objects in the 2014 epoch observations.}

\begin{tabular}{llccrrrrc}
\hline
\textbf{Name} & \textbf{Comp.} & \boldmath$\mathrm{S_6}$ & \boldmath$\mathrm{S_9}$ & \boldmath$\alpha$ & \boldmath$\theta^6_{maj}\;\;\;\;$ & \boldmath$\theta^9_{maj}\;\;\;\;$ & \boldmath$\gamma\;\;\;\;\;\;\;$ & \textbf{R}\\
 & & ($\uJy$) & ($\uJy$) & & \centering$(\arcsec)\;\;\;\;\;\;\;$ & $(\arcsec)\;\;\;\;\;\;\;$ & \\
\hline
G263.7434+00.0161 & N & $626\pm43$ & $594\pm44$ & $-0.13\pm0.25$ & $1.40\pm0.19$ & $1.40\pm0.14$ & $-0.01\pm0.41$ & 0.5 \\
 & S2 & $319\pm28$ & $363\pm31$ & $0.32\pm0.30$ & $1.13\pm0.49$ & $1.04\pm0.27$ & $-0.20\pm1.25$ & 0.5 \\
 & S & $767\pm52$ & $837\pm58$ & $0.22\pm0.24$ & $2.24\pm0.13$ & $1.85\pm0.10$ & $-0.47\pm0.19$ & 0.5 \\
 & SE & $126\pm22$ & $135\pm29$ & $0.17\pm0.68$ & $5.18\pm1.09$ & $5.29\pm1.27$ & $0.05\pm0.79$ & 0.5 \\
G310.0135+00.3892 & N & $707\pm53$ & $695\pm66$ & $-0.04\pm0.30$ & $0.28\pm0.07$ & $-$ & $-$ & -1 \\
 & S & $347\pm37$ & $481\pm53$ & $0.81\pm0.38$ & $0.50\pm0.16$ & $-$ & $-$ & -1 \\
 & SW & $126\pm35$ & $93\pm32$ & $-0.75\pm1.09$ & $<2.60$ & $-$ & $-$ & -1 \\
G310.1420+00.7583A & A1 & \multirow{ 2}{*}{\Big\}$2744\pm177$} & $1459\pm122$ & $-$ & \multirow{ 2}{*}{\Big\}$0.84\pm0.03$} & $0.66\pm0.28$ & $-$ & -1 \\
 & A2 & & $1852\pm147$ & $-$ & & $0.62\pm0.06$ & $-$ & -1 \\
 & A3 & $2053\pm144$ & $1295\pm160$ & $-1.14\pm0.35$ & $1.09\pm0.04$ & $1.36\pm0.14$ & $0.53\pm0.28$ & -1 \\
 & A4 & $888\pm83$ & $570\pm94$ & $-1.09\pm0.47$ & $0.93\pm0.10$ & $1.00\pm0.22$ & $0.18\pm0.60$ & -1 \\
 & B & $205\pm65$ & $117\pm61$ & $-1.38\pm1.50$ & $2.78\pm0.94$ & $-$ & $-$ & -1 \\
 & C & $568\pm56$ & $511\pm74$ & $-0.26\pm0.43$ & $<0.51$ & $0.50\pm0.17$ & $>-0.03\pm0.82$ & -1 \\
 & D & $1896\pm137$ & $1684\pm157$ & $-0.29\pm0.29$ & $1.09\pm0.05$ & $0.98\pm0.08$ & $-0.25\pm0.22$ & -1 \\
G313.7654$-$00.8620 & A1 & $506\pm46$ & $537\pm55$ & $0.15\pm0.34$ & $<0.57$ & $0.43\pm0.10$ & $>-0.67\pm0.60$ & -1 \\
 & A2 & $183\pm39$ & $170\pm60$ & $-0.18\pm1.02$ & $<3.00$ & $1.92\pm0.76$ & $>-1.10\pm0.98$ & -1 \\
 & B1 & $350\pm54$ & $240\pm76$ & $-0.93\pm0.87$ & $1.74\pm0.25$ & $1.92\pm0.62$ & $0.24\pm0.87$ & -1 \\
 & B2 & $116\pm27$ & $167\pm46$ & $0.90\pm0.89$ & $-$ & $1.20\pm0.41$ & $-$ & -1\\
 & B3 & $118\pm30$ & $81\pm27$  & $-0.93\pm1.03$& $2.15\pm0.90$ & $1.97\pm1.11$ & $-0.22 \pm 1.73$ & 0.5\\
 & C & $420\pm44$ & $399\pm57$ & $-0.13\pm0.44$ & $0.85\pm0.12$ & $1.06\pm0.15$ & $0.53\pm0.50$ & -1\\
 & D & $138\pm25$ & $157\pm28$ & $0.32\pm0.62$ & $-$ & $-$ & $-$ & -1\\
 & F & $85\pm22$ & $85\pm25$ & $0.00\pm0.97$ & $<2.2$ & $-$ & $-$ & 0.5\\
 & G & $171\pm59$ & $51\pm18$ & $-2.98\pm1.02$ & $4.75\pm1.56$ & $-$ & $-$ & 0.5\\ 
\hline
\end{tabular}
\label{tab:JetsMainProperties}
\end{table*}

\subsection{G263.7434+00.1161}
\label{sec:g263.7434}
G263.7434+00.1161 is the nearest and lowest-luminosity object in our sample, located in the Vela molecular ridge cloud D at a distance of $0.7\pc$. Observations at 1.2mm by \citet{Massi2007} revealed an $18\Msol$ compact core (MMS12, $0.13$ pc in size) peaking $\sim5\arcsec$ to the east of our pointing centre. This mm-core is coincident with `complex and intense' H2 emission at $2.12\micron$ \citep{deLuca2007}, indicative of collisional excitation of molecular hydrogen through shocks attributable to protostellar outflows \citep{Wolfire1991}. \citetalias{Purser2016} identified two radio sources, one centred on the MYSO's position (S) and the other offset $\sim4\arcsec$ to the NW. Derived spectral indices were thermal ($\alpha=0.4\pm0.2$) and non-thermal ($\alpha=-0.5\pm0.3$) for S and N respectively.

\autoref{fig:g263_7434contour} shows this work's maps of radio flux towards G263.7434+00.1161. Three lobes of emission, N, S and SE, are detected at both frequencies and aligned along a position angle of $\sim158\degr$. Most obviously in the more uniformly-weighted images (panels c and d of \autoref{fig:g263_7434contour}), the thermal jet (S) is in fact elongated along a different axis to that running through all three components (see \autoref{sec:precession} for further discussion of this). Component SE, whose peak is located $\sim4\arcsec$ from S at a position angle of $\sim160\degr$, is clearly elongated along a position angle of $178\degr$ with a length of $\sim5\arcsec$ ($3500\au$ at a distance of $0.7\pc$) and is calculated to possess a spectral index of $0.17\pm0.68$. Approximately $29\arcsec$ to the south ($\alpha_\mathrm{J2000} = 08\rahr48\ramin48.63\rasec$, $\delta_\mathrm{J2000} = -43\degr32\arcmin57.8\arcsec$) we detect a new source in both the $6$ and $9\GHz$ clean maps which does not appear in the 2013 epoch's images (see \autoref{fig:g263_7434_Radio_Star} of the Appendix for a contour plot of radio flux), which we designate S2. With integrated fluxes of $319\pm28$ and $363\pm31\uJy$ at $6$ and $9\GHz$, it possesses a thermal spectral index of $\alpha=0.32\pm0.30$ and has a GLIMPSE, mid-IR counterpart. 


\subsection{G310.0135+00.3892}
\label{sec:g310.0135}
G310.0135+00.3892 (or IRAS 13481$-$6124) is well-studied across a broad range of wavelengths. \citet{Kraus2010} used NIR interferometric observations ($\theta_\mathrm{res} = 2.4\,\mathrm{mas}$ or $8.4\au$ at $\theta_\mathrm{PA} = 114\degr$) to directly observe a hot, dusty, compact ($13\au \times 19\au$) disc. \citet{Ilee2013} detected CO bandhead emission and subsequent modelling found temperature gradients consistent with the findings of \citet{Kraus2010} and with a flared, irradiated disc around a $ 21.8\Msol$ MYSO. The disc's major axis orientation was found to be perpendicular to a collimated, (opening angle $\sim6\degr$) CO, bipolar outflow \citep[red lobe to the NE,][]{Kraus2010}, itself aligned with two, $4.5\,\mu\mathrm{m}$ excesses (from inspection of IRAC imagery) separated by $6.5$ pc, indicative of outflow activity. Corroborating these findings, diffraction limited MIR imaging at $20\micron$ by \citet{Wheelwright2012} suggested that the dominant emission at $20\micron$ was from the walls of cavities evacuated by outflows along a NE-SW axis. \citet{Caratti2015} observed the $\mathrm{H}_2\, 2.122 \micron$ transition, detecting lobes of emission spread over $6.9\pc$ at a position angle of $-154\degr$, parallel with the established molecular outflow. These lobes appeared to be more spread out to the NE (red lobe) showing that a density gradient exists in this direction. \citetalias{Purser2016} found 3 radio lobes associated with the MYSO designated N, S and SW with spectral indices of $-0.2\pm0.1,1.3\pm0.2$ and $0.7\pm1.7$ respectively. While the S and SW lobes are aligned to the general outflow direction, N is offset by $\sim30\degr$ to the west. It was deduced that while S represents the MYSO and base of the jet, the SW component is the faint thermal emission from the jet itself while N is likely optically thin and/or non-thermal emission as the result of wide-angle shocks from the jet on the surrounding material. The reason for the offset was unclear, however \citet{Caratti2016} observed spatially, and spectrally, resolved Br$\gamma$ emission whose spatial velocity profile (their Figure 2) seemed to show a wide range in outflow angle, for a bipolar ionized jet with a terminal velocity of $500\,\mathrm{km\,s^{-1}}$. This could possibly account for shock sites significantly deviating from the jet's outflow angle \citep[$26\degr$ according to][]{Caratti2015}. 


In the clean maps presented in \autoref{fig:g310_0135contour}, the previously established N, S and SW components are detected at both frequencies. The SW lobe appears to `break up' into two separate components from $6$ to $9\GHz$ (comparing panels a and b of \autoref{fig:g310_0135contour}), separated by $1.4\arcsec$. However this may be due to image defects caused by strong, residual, sidelobes from a $\sim27\mJy$ source $\sim470\arcsec$ to the NW of the pointing centre. Imaging the whole of the primary beam at $6\GHz$ also shows three extended lobes of emission, designated as HH1, HH2 and HH3 (shown in \autoref{fig:g310_0135_HH}, with positions recorded in \autoref{tab:JetsLobesPositions6GHz} of appendix \ref{app:tables}). These sources lie at separations of $0.88,\,2.47$ and $3.78\pc$ with position angles of $30, 27$ and $31\degr$ from the MYSO respectively. No spectral indices could be computed since $9\GHz$ clean maps resolved out much of the extended emission of these lobes. 


\subsection{G310.1420+00.7583A}
\label{sec:g310.1420}
G310.1420+00.7583A is associated with IRAS 13484--6100 and is offset by $\sim8\arcsec$ from a cometary UCHII (G310.1420+00.7583B). The radio observations of \citetalias{Purser2016} detected 7 separate components, named A1 (MYSO), A2, A3, A4, B, C and D, roughly aligned east to west with a total, integrated flux of $\sim10\mJy$. Previous to that \citet{Urquhart2007ATCA} detected radio emission (observations conducted in November 2004) at 4.8 GHz coincident with components A1 and A2, with an integrated flux of $2.92\pm0.75\mJy$ (fitted from the archived image using \textsc{imfit}). Different masing species have been detected towards this source \citep[OH, H$_2$O and CH$_3$OH:][respectively]{Walsh1998, Urquhart2009, Green2012}, with the methanol maser coincident (multibeam survey positional accuracy is $<0.1\arcsec$) with A1. \citet{Caratti2015} observed two knots of H$_2$ emission, one of which also displayed Br$\gamma$ emission indicative of strong dissociative shocks (their `knot 1' or D of \citetalias{Purser2016}) with shock velocities $>90\kmps$ within a medium of density $\sim10^5\,\cmc$. Jet properties inferred from the H$_2$ observations included a length of $0.4\pc$, precession of $17\degr$ and electron density, $n_e$, of $(4\pm1)\times10^4\,\cmc$. An extended green object was also detected offset from knot 1 by $\sim16\arcsec$ at a PA of $57\degr$ \citep{Cyganowski2008}. 

The clean maps shown in \autoref{fig:g310_1420contour} show all components detected by \citetalias{Purser2016} aligned in a jet-like morphology along a position angle of $\sim78\degr$. One previously-undetected component is seen at the $6\sigma$ level (panel a), located $\sim5.5\arcsec$ to the east of component D. At $9\GHz$ it is not detected and an upper limit to the spectral index of $\alpha<0$ is derived from the peak flux at $6\GHz$. We also detect radio emission roughly connecting lobe C to D, which is especially well shown in panels b and c of \autoref{fig:g310_1420contour}, and methanol maser emission coincident with A1. Fluxes and positions for A1, A2, A3 and A4 proved difficult to measure via \textsc{imfit} due to source confusion. Using clean maps with a robustness of 0.5, the emission from A1 and A2 could not be separated and only at $9\GHz$, using a robustness of $-1$, could we deconvolve each individually. For components B, C and D this was not the case and consequently sizes and integrated fluxes are derived from the clean map with a robustness of 0.5 to maximise the signal to noise ratio. A resolved out \textsc{Hii} region is also seen to the south west of A1 (RMS survey alias G310.1420+00.7583B or component E from \citetalias{Purser2016}) which is not discussed further.

\subsection{G313.7654$-$00.8620}
\label{sec:g313.7654}
Associated to IRAS 14212$-$6131, the observations of \citetalias{Purser2016} detected 6 associated radio components named A1 (MYSO), A2, B1, B2, C and D. Both hydroxyl \citep{Caswell1998} and methanol \citep{Green2012} masers have been previously observed, the latter of which was separated from A1 by $\sim1\arcsec$ at a position angle of $108\degr$. \citet{Caratti2015} detected 4 knots of H$_2$ emission which, if tracing a jet, show evidence of a precession in the jet's axis of $32\degr$. Knot 1 (coincident with B1 of \citetalias{Purser2016}) has an inferred electron density of $(1\pm0.5)\times10^4 \cm^3$, while knot 4 (most distant) displays Br$\gamma$ emission indicative of strong J-type shocks with a shock speed of $\sim60\kmps$. Currently the jet axis is defined at a position angle of 125$\degr$, with a length (on one side) of $1.4\pc$. GLIMPSE images show diffuse $4.5\micron$ excesses in the general area of previously established radio emission, an extended \HII region $\sim30\arcsec$ to the west and a compact \HII region $\sim10\arcsec$ to the south west \citepalias[E from][]{Purser2016}.

Two new components are seen in the clean maps of \autoref{fig:g313_7654contour} compared to  \citetalias{Purser2016}. One is situated $\sim2\arcsec$ to the NW of B1 which we designate as B3, and the other approximately halfway between A2 and C, which is relatively diffuse and is therefore unnamed. For B3 we derive a spectral index of $\alpha=-0.9\pm1.0$. Outside the clean maps of \autoref{fig:g313_7654contour} a radio source (shown in \autoref{fig:g313_7654_F} in the online version) is detected $\sim24\arcsec$ to the NE of A1, named F, for which we measure integrated fluxes of $85\pm18\uJy$ and $85\pm21\uJy$ at 6 and 9 GHz respectively and derive $\alpha=0.0\pm1.0$. It is also interesting to note that there is an extended radio lobe with spectral index $\alpha=-3\pm1$ (suffering from resolving-out effects at $9\GHz$) detected $\sim33\arcsec$ from A1 at $\theta_\mathrm{PA}\sim-81\degr$, which we name G. Two 6.7 GHz methanol maser spots are also detected, one coincident with A1 (see \autoref{fig:g313_7654contour}) and the other coincident with F. For the calculation of spectral indices, a robustness of $-1$ was employed at all frequencies to allow for the effective deconvolution and subsequent measurement of all component's (apart from B3) integrated fluxes. Component B3 was only detected in the clean maps with a robustness of 0.5 but was sufficiently separated from B1/B2 to allow the accurate deconvolution and measurement of its flux. For all lobes, calculated spectral indices agree with the values obtained in \citetalias{Purser2016}. 



\section{Necessary considerations}
\label{sec:considerations}
Prior to the analyses conducted in \autoref{sec:discussion}, two issues affecting the direct comparison of images from different epochs required careful consideration. In the rest of this section we highlight their implications upon our analyses, and our approach to negate, or at least accommodate for, their effects.

\subsection{Image alignment}
\label{sec:imagealignment}
Measurement of the positional changes of radio lobes over subsequent observations allows for the direct measurement of the ionized gas' velocity (in the plane of the sky). In order to do this, accurate positional measurements are required due to the great distances involved, particularly for our sample which has relatively short time baselines. For example with two measurements separated by a period of $2\yr$, assuming a jet velocity of $500\kmps$ for a distance of $3\kpc$, we expect an angular shift (assuming the jet's path lies in the plane of the sky) of $0.07\arcsec$. However due to imperfections in calibration, as well as astrometric inaccuracy, coordinates have an absolute positional uncertainty. In an effort to negate this effect, we used the same phase calibrators for our new observations as for those of 2013. Extragalactic background sources were subsequently identified from the clean maps of radio flux (on the basis of spectral index and lack of IR counterparts) and any positional shift of these objects between the two images was assumed to be due to positional errors. This provided corrections which were applied to the 2014 images using spline interpolation, therefore aligning the two images with subsequent analyses utilising these `corrected' clean maps.

\subsection{Variable ATCA flux recovery}
\label{sec:atcafluxrecovery}
\begin{figure*}
\includegraphics[]{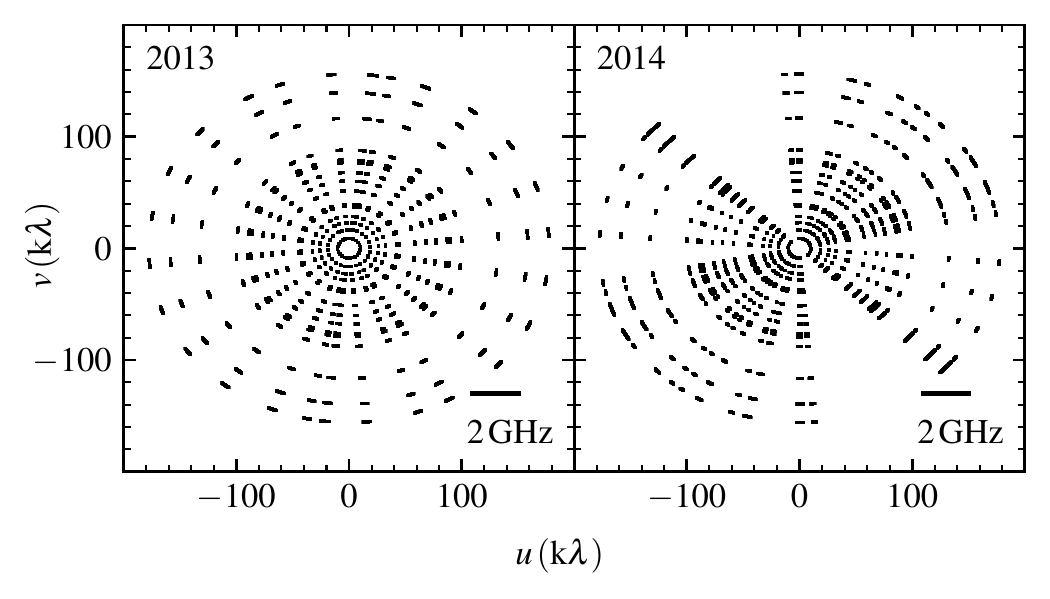}
\caption{Plots of the sampling functions (single channels at 9 GHz) for the 2013 and 2014 datasets towards G310.1420+00.7583A. Illustrated in the bottom right of each plot is a scale bar showing the radial uv coverage over the 2 GHz bandwidth ($6\km$ baseline).}
\label{fig:uvcoverages}
\end{figure*}

Different sampling of the $uv$-plane for each of the two epochs can change recovered flux densities, distributions and morphologies. To investigate the effects this difference has on the resulting clean maps, a set of synthetic observations were created mirroring both the system temperatures and sampling functions, $S(u,v)$, of the 2013 and 2014 data (shown in \autoref{fig:uvcoverages}) taken towards one of our targets (G310.1420+00.7583A). This particular object was chosen on the basis of its complex morphology which contains both separate and spatially-confused emission. The model chosen was one identical to the (\textsc{imfit}-derived) positions, integrated fluxes and sizes recorded in the 2013 data of the 6 most prominent lobes (A1, A2, A3, A4, C and D). Resulting clean maps and pixel-to-pixel flux difference maps are presented in Figures \ref{fig:SOcleanmaps} and \ref{fig:SOfluxdiff} respectively of appendix \ref{app:figures} (online version only). The subsequent analysis of this synthetic data showed that relative gaps in $uv$-coverage have acted to decrease the amount of recovered flux from 2013 to 2014 by $-16\%$. Predictably this effect is amplified for the more extended components, C and D, but the negative percentage change is apparently greater for A3 (a relatively compact lobe). Further to this, the integrated flux actually increases for A2 in both epochs. Both of these effects however are not solely due to the differing synthetic apertures, but also with the inaccuracy of fitting multiple, closely-spaced, Gaussian distributions of flux. This is reflected again in the inability of the 2014 synthetic observations to deconvolve A2 as a point source (as in the model), and subsequently it was measured to have finite angular size and position angle. Component A4 also appears to significantly relocate by $\sim0.19\arcsec$, at a position angle of $-16\degr$, for the 2014 dataset compared to the model. This is equivalent to a proper motion derived velocity of $\sim2900\kmps$ for G310.1420+00.7583A's distance of $5.4\kpc$. In summary we can say the following:

\begin{itemize}
\item Fluxes recovered by the synthetic observations are on average $-11\%$ and $-25\%$ lower than actual values supplied by the model, with extended sources C and D affected worst.
\item Percentage of flux recovered for the 2014 observations is lower than for the 2013 observations, whereby, on average, $16\%$ less flux is recovered per component.
\item Source confusion leads to the wrong distribution of flux for lobes A2 and A3, with the compact source A2 `leeching' flux from A3.
\item Lobe positions are deduced with better than $0.1\arcsec$ accuracy in non-source confused components.
\item Position angles for the emission are relatively well recovered by \textsc{imfit}, with average deviations of just $-2\degr$ and $6\degr$ (ignoring source confused lobes, A2 and A3) for the former and latter epochs respectively.
\end{itemize}

This analysis implies that for complex sources such as G310.1420+00.7583A and G313.7654$-$00.8620, any proper motions $<0.1\arcsec$ are likely too affected by poor uv-coverage and source confusion to draw any conclusions from. Negative variations over time in integrated lobe fluxes between 2013 and 2014 should be regarded with caution. Pixel-to-pixel comparisons of the maps of flux also show non-real negative flux variations centred on components A1, A2, A3 and A4. These considerations are therefore necessarily taken into account for the rest of this work.

\section{Discussion}
\label{sec:discussion}
Observations presented in \autoref{sec:results} have detected new radio emission towards a small sample of young stellar objects across a range in bolometric luminosity. We now move on to discuss our results and their implications upon the natures of newly detected faint emission, the variability/proper motions of the jets and their lobes and precession in the jets' outflow axes.

\subsection{Natures of newly detected emission}
\label{sec:faintemission}
Towards all of the objects in the sample, apart from G310.1420+00.7583A, sources of radio emission below the detection threshold of previous observations are detected, resulting from many different processes.

\begin{figure}
\centering
\includegraphics[width=\columnwidth]{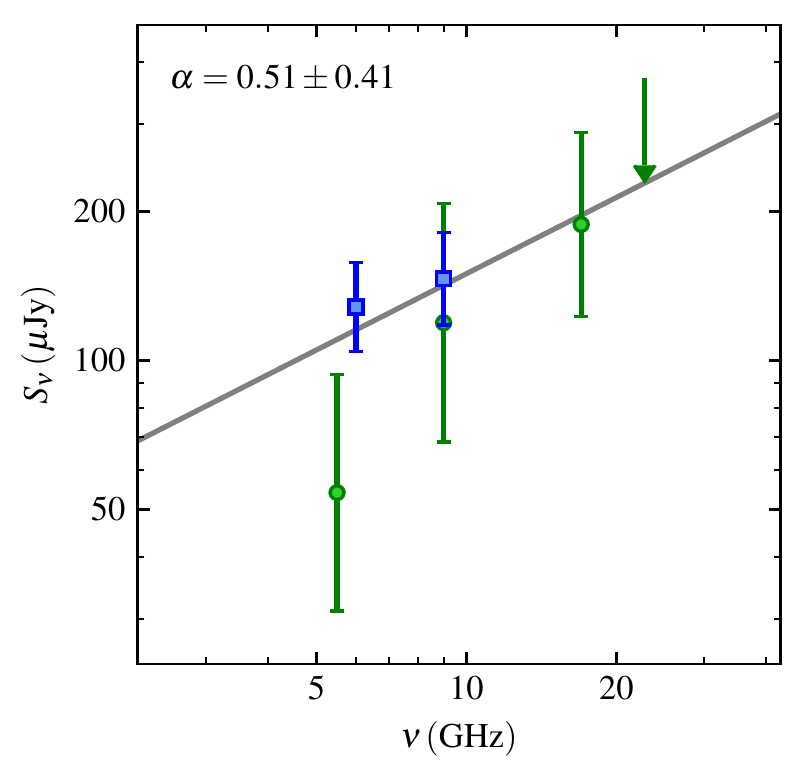} 
\caption{A plot of integrated flux against frequency for the SE component of G263.7434+00.1161. Green circles and blue squares represent data from \citetalias{Purser2016} and this work respectively, with the upper limit being that of the $22.8\GHz$ data from \citetalias{Purser2016}. All error bars shown include a $5\%$ uncertainty in the absolute flux calibration. The grey line shows the (log-space) weighted, least-squares fit of the combined data from both epochs, with the derived spectral index shown in the top-left corner.}
\label{fig:g263spectrum}
\end{figure}

In the case of G263.7434+00.1161's SE component, morphology of the emission at $6\GHz$ shows it to be extended in a jet-like, elongated structure $(5.2\times0.6)\arcsec$ at a position angle of $179\degr$. However the spectral index derived from $6$ to $9\GHz$ ($\alpha=0.17\pm0.68$) was unable to establish whether is was thermal or non-thermal in nature. In an attempt to constrain the emission processes at work, the data from \citetalias{Purser2016} were re-imaged with a robustness of 2 (in order to maximise sensitivity) and a $3\sigma$ component was detected, coincident with the new SE lobe, at $5.5, 9$ and $17\GHz$. Least square fitting to the first epoch's data alone yields a spectral index of $1.06\pm0.61$, while for the combined fluxes of all available data a spectral index of $\alpha=0.51\pm0.41$ is calculated (see \autoref{fig:g263spectrum}). This value for the spectral index ($0.51\pm0.41$) determines it to be the direct, thermal emission from the jet's stream and is compatible with a wide variety of scenarios from the models of \citet{Reynolds1986}, including the `standard' spherical case of a freely flowing ionized jet. Considering its properties and alignment with the North-South component axis, this likely represents the fainter, direct, thermal emission from the jet's stream. A discussion of the new radio object, S2, detected $29\arcsec$ to the south of G263.7434+00.1161 is reserved for \autoref{sec:var+pms}.

With G310.0135+00.3892, in comparison to other objects in this sample, the previously undetected emission (HH1, HH2 and HH3) is found to be greatly separated from the MYSO itself (up to $3.78\pc$) but still roughly aligned along the same axis of $\sim30\degr$ with respect to S, much like the HH80-81 system \citep[which has Herbig-Haro objects spread over $5.3\pc$,][]{Marti1993}. In comparison with a previous near-infrared study \citep{Caratti2015}, HH1, HH2 and HH3 are spatially coincident with their $2.122\micron$, H$_2$ lobes `E red', `D red' and `bow-shock A Red'. Considering that this type of emission is the result of shocks with velocities $>15\kmps$ \citep{Elias1980}, this confirms HH1, HH2 and HH3's status as radio Herbig-Haro lobes. From their integrated fluxes and sizes, average emission measures of $5350 \pm 825, 9590 \pm 1250$ and $9990 \pm 2110\, \rm pc\,cm^{-6}$ are calculated, while average electron densities are found to be $303 \pm 55, 351 \pm 61$ and $549 \pm 116\, \cm^{-3}$ for HH1, HH2 and HH3 respectively. 

An important point for discussion is the relatively asymmetric nature of the radio flux's distribution towards G310.0135+00.3892, whereby the SW component has no corresponding NE lobe. We know from previous observations that a bipolar, collimated, ionised jet is present at both large and small scales \citep[][respectively; see \autoref{sec:g310.0135}]{Caratti2015,Caratti2016} and therefore a monopolar jet is unlikely. With uncertain spectral indices, the SW component may arise from a variety of physical mechanisms. If the SW lobe was shocked emission from the jet impinging upon clump material, it could be a reflection of the asymmetry of its environment. Indeed, the distribution of H$_2$ shock emission seen by \citet{Caratti2015} was postulated to be due to a higher density of material towards the SW, in comparison to the NE, a picture supported by the MYSOs offset of $7\arcsec$ to the NE of the peak position of its parental clump \citep{Contreras2013}. Alternatively, the presence of the accelerating flow \citep{Caratti2016} suggests largely evacuated outflow cavities and therefore the shock site may reside in the cavity walls. Finally, if SW is a shock internal to the jet, or indeed thermal emission from the jet's stream, this could be due to asymmetric ejection in velocity and/or mass flux. Only further radio observations over longer time baselines could discern between these potential explanations.

\begin{figure}
\centering
\includegraphics[width=\columnwidth]{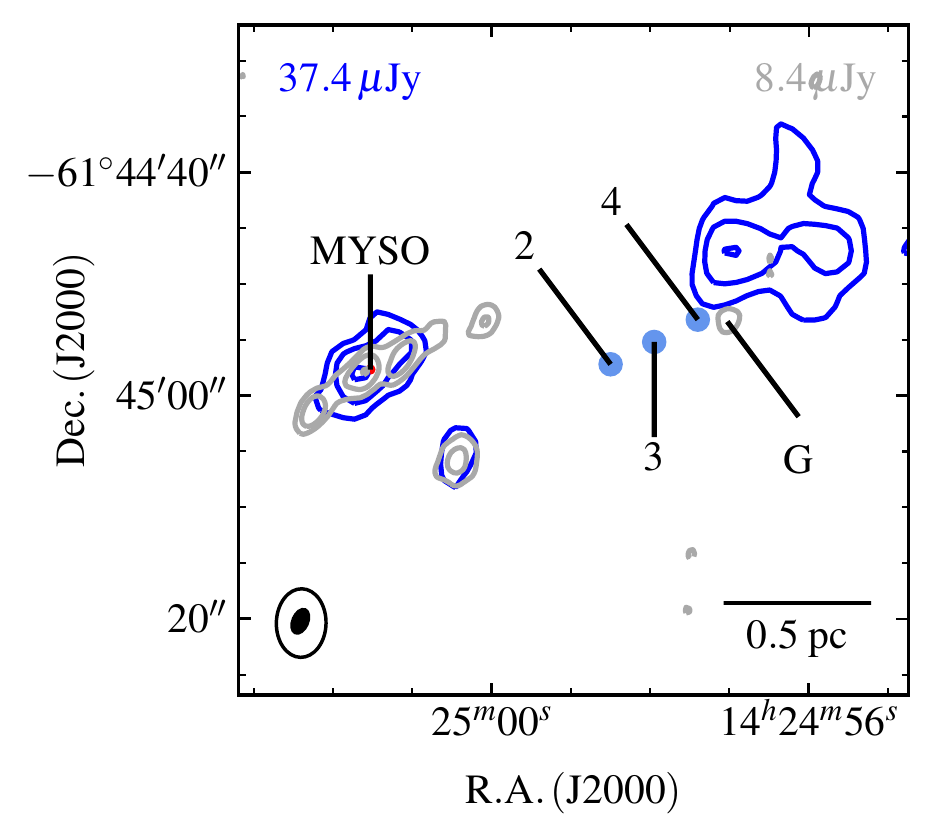}
\caption{Radio contour map of the general area towards G313.7654$-$00.8620 utilising a robustness of $0.5$, at $2\GHz$ (blue contours) and $6\GHz$ (grey contours). Restoring beams are illustrated in the bottom left corner and the $2\GHz$ beam has dimensions of $6.14\arcsec \times 4.45\arcsec$ at $\theta_{PA}=-1.24\degr$. Contours are set at $(-4, 4, 6, 10)\times\sigma$ and $(-4, 4, 15, 58)\times\sigma$ for the $2$ and $6\GHz$ data respectively. The value employed for $\sigma$ is indicated in the top-left corner (for $2\GHz$) and top-right (for $6\GHz$). Filled, blue circles indicate the $2.12\micron$ knots, `knot 2', `knot 3' and `knot 4' from \citet{Caratti2015}. The radio lobe detected at $6\GHz$, which suggests an overall precession, in the outflow axis is annotated.}
\label{fig:g313_7654_2GHz}
\end{figure}

Diffuse faint emission between the previously established lobes, A2 and C of G313.7654$-$00.8620 was detected, which is likely `filling in' the jet's emission as a result of the increased sensitivity of this set of observations.  Component F on the other hand is distinctly separated from the main radio emission, compact and has a methanol maser detected towards it meaning it is likely another MYSO in the vicinity. However it has no clear near or mid-IR counterpart and therefore must be deeply embedded within the clump and/or relatively unevolved. Maps of $2.122\,\micron$ ${\rm H_2}$ emission seen by \citet{Caratti2015} shows the other new radio component B3 to be coincident with their `knot 1', a fact which, in conjunction with B3's alignment to the presumed jet outflow axis, classifies it as a likely jet-shocked surface. Unfortunately B3's spectral index was calculated to be $-0.9\pm1.0$ meaning the exact emission processes at work could not be constrained. Further from the obvious string of jet-lobes is radio lobe G detected $33\arcsec$ to the west. Considering the relative position angle of the emission from the thermal jet (A1) and coincidence with a NIR H$_2$, 2.122$\micron$ lobe \citep[knot 4 of][]{Caratti2015}, we propose this to be another surface upon which the jet of outflowing material is impinging. This position also sits at the head of a cometary \textsc{Hii} region's bow-shock, from which extended, partially resolved out emission is detected at $6\GHz$. It is therefore possible that this new radio emission is attributable to the \textsc{Hii} region and not with a shocked surface along the jet's axis. However archival observations, taken on 09/12/2011 at $2\GHz$, show the \textsc{Hii} region's emission to be offset to the detected radio lobe (\autoref{fig:g313_7654_2GHz}), and extended along a position angle of $\sim95\degr$ across an extent of $\sim10\arcsec$ (perpendicular to the MIR \textsc{Hii} region's `front').  Since the \textsc{Hii} region at $2\GHz$ is not co-located with the radio lobe at $6\GHz$, which itself is coincident with NIR shock emission, this lobe may still be attributable to jet activity.

\subsection{Variability and proper motions}
\label{sec:var+pms}
To examine both the flux variability and motion of radio sources, a comparison of both the \textsc{imfit}-derived lobe positions and integrated $9\GHz$ fluxes from \citetalias{Purser2016} with those recorded here was performed (tabulated in \autoref{tab:JetsLobesPMsVars}). Considering the short time-baseline between the two datasets, the $3\sigma$ lower-limits (assuming $\sim0.1\arcsec$ positional uncertainty) on the proper motions are relatively high, ranging from $600-6000\kmps$ and therefore realistically their detection was not expected.


\begin{table*}
\caption{A table of the proper motion-derived velocities/position angles and integrated flux changes from the first to second epochs. These quantities are calculated from the \textsc{imfit}-derived positions/fluxes at $9\GHz$. Upper limits given are the $3\sigma$ upper limits. Velocities incorporate errors on distances of $1\kpc$ (i.e. spiral arm width) apart from for G263.7434+00.1161 whose distance error is $0.2\kpc$ \citep{Liseau1992}.}
\label{tab:JetsLobesPMsVars}
\begin{tabular}{lccccccccc}
\toprule 
\textbf{Source} & \textbf{Lobe} & \boldmath$r$ & \boldmath$v_\mathrm{PM}$ & \boldmath$\theta$ & \boldmath$\Delta S_9$ & \textbf{R} \\
 & & $(\arcsec)$ & $(\kmps)$ & $(\degr)$ & $(\uJy)$ &  \\
\midrule
G263.7434+00.1161 & N & $0.032\pm0.062$ & $<339$ & $29\pm66$ & $<270$ & $0.5$ \\
 & S & $0.060\pm0.032$ & $<183$ & $245\pm63$ & $<375$ & $0.5$ \\
 & SE & $0.716\pm0.388$ & $<2208$ & $241\pm54$ & $<276$ & $0.5$ \\
 & S2 & $-$ & $-$ & $-$ & $>283\pm41$ & $0.5$\\
G310.0135+00.3892 & N & $0.101\pm0.035$ & $<1188$ & $307\pm24$ & $<324$ & $-1$ \\
 & S & $0.071\pm0.052$ & $<1431$ & $346\pm24$ & $<276$ & $-1$ \\
 & SW & $0.092\pm0.140$ & $<3582$ & $338\pm42$ & $<219$ & $-1$ \\
G310.1420+00.7583A & A1 & $0.097\pm0.012$ & $1376\pm308$ & $97\pm26$ & $<591$ & $-1$ \\
 & A2 & $0.115\pm0.036$ & $<1791$ & $35\pm13$ & $933\pm171$ & $-1$ \\
 & A3 & $0.238\pm0.043$ & $3371\pm867$ & $309\pm12$ & $<726$ & $-1$ \\
 & A4 & $0.453\pm0.079$ & $6409\pm1627$ & $351\pm3$ & $<505$ & $-1$ \\
 & B & $0.318\pm0.121$ & $<5718$ & $186\pm8$ & $<285$ & 0.5 \\
 & C & $0.128\pm0.035$ & $1806\pm596$ & $70\pm26$ & $<552$ & 0.5 \\
 & D & $0.068\pm0.049$ & $>2145$ & $344\pm16$ & $<699$ & 0.5 \\
G313.7654$-$00.8620 & A1 & $0.063\pm0.020$ & $<1320$ & $244\pm27$ & $<321$ & $-1$ \\
 & A2 & $0.754\pm0.149$ & $15404\pm3631$ & $309\pm13$ & $304\pm82$ & 0.5 \\
 & B1 & $0.196\pm0.109$ & $<6825$ & $195\pm17$ & $<225$ & 0.5 \\
 & B2 & $0.511\pm0.100$ & $10444\pm2449$ & $327\pm8$ & $<135$ & 0.5 \\
 & C & $0.043\pm0.104$ & $<6363$ & $161\pm68$ & $<372$ & $-1$ \\
 & D & $0.110\pm0.052$ & $<3279$ & $326\pm24$ & $<258$ & $-1$ \\
\bottomrule
\end{tabular}
\end{table*} 

\begin{figure}
\centering
\includegraphics[width=\columnwidth]{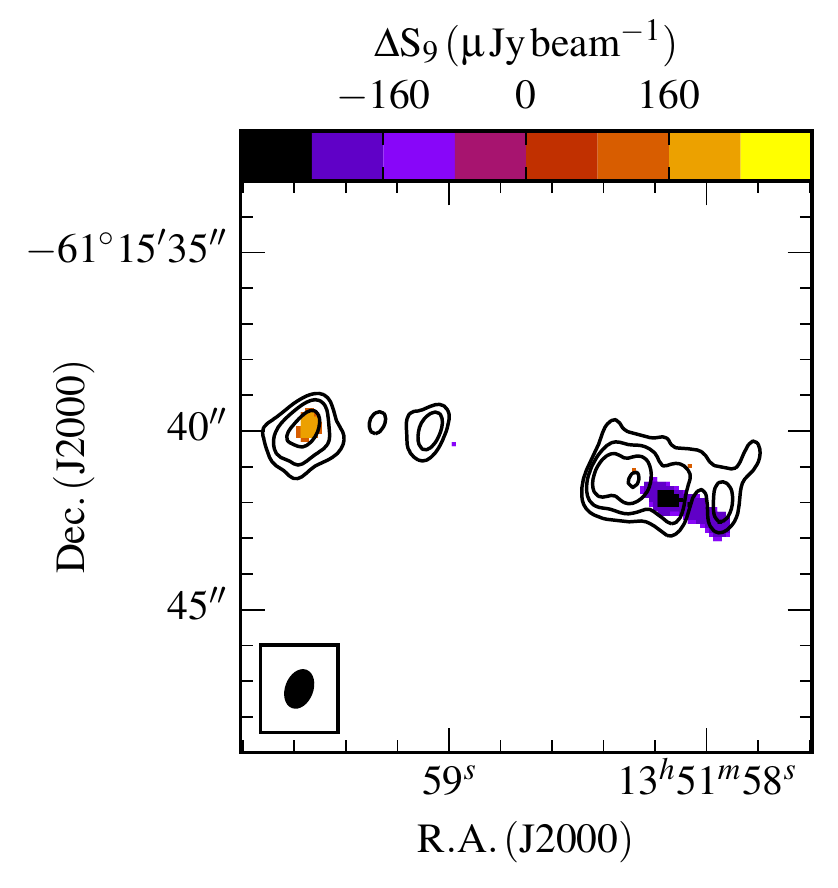} 
\caption{A plot of the absolute flux difference between the 2014 and 2013 $9\GHz$ clean maps (robustness = -1) of radio flux for G310.1420+00.7583A (colourscale). Only the pixels whose flux difference was $\Delta S_\nu\leq-4\sigma\,|\,\Delta S_\nu\geq4\sigma$ were included (where $\sigma=34.5\uJy\,\mathrm{beam}^{-1}$, the rms noise in the map of absolute flux difference). The common restoring beam's dimensions were $1.05\arcsec\times0.69\arcsec$ at $\theta_\mathrm{PA}=-20.9\degr$. Contours represent the $9\GHz$ data taken in 2014 with $-4, 4, 9, 21$ and $47\sigma$ shown ($\sigma = 22.4\uJy\,\mathrm{beam}^{-1}$).}
\label{fig:g310_1420fd}
\end{figure}

From the \textsc{imfit}-measured 2013 and 2014 $9\GHz$ fluxes, only two components directly attributable to the jets show significant ($>3\sigma$) variability, being component A2 of G310.1420+00.7583A and A2 of G313.7654$-$00.8620. However, since many of the sources' flux distributions (especially these two) are not simple, it is unclear whether this change is due to inaccuracies in either deconvolution, measurement of the emission using \textsc{imfit}, or both. A pixel-to-pixel comparison and analysis helped to reveal that variability is likely present towards G310.1420+00.7583A, without relying on accurate deconvolution by a multiple set of Gaussians in the image plane. In \autoref{fig:g310_1420fd}, the absolute change in the flux for each pixel is plotted for G310.1420+00.7583A (\autoref{fig:g310_1420fd}) showing well-ordered variability towards the western side of the jet (i.e. A2, A3 and A4). However, as discussed in \autoref{sec:atcafluxrecovery}, this was an expected effect of differing sampling functions for the two sets of observations (comparing \autoref{fig:g310_1420fd} to its synthetic equivalent in \autoref{fig:SOfluxdiff} of appendix \ref{app:figures}). Unexpectedly from considerations of $uv$-coverages, component D shows an increase in the pixel fluxes towards its peak flux position, not reflected in its integrated flux (via measurement with \textsc{imfit} or integrating flux within the $3\sigma$ contours). Therefore we suggest that D is a spatially-resolved shock-site which is evolving over time.

Away from the obvious emission along the ionized jets' axes, a highly variable radio source (S2) was detected $29\arcsec$ to the south of G263.7434+00.1161 which should have been seen at the $\sim12\sigma$ level in the 2013 data and as such possesses a lower-limit on its $9\GHz$ flux change of $>283\pm41\uJy$. The thermal nature of the radio emission precludes extragalactic origins since only starburst galaxies rich in \textsc{Hii} regions should display thermal spectral indices, but their fluxes should not be variable on the time-scale of $\sim2\yr$. At a position angle of $181\degr$ from S, which is roughly aligned with the deconvolved position angle of S's major axis, it is possible that this source may be a highly variable radio Herbig-Haro object. On the other hand, it is separated by $0.17\arcsec$ (which is likely coincident within astrometric errors) from a 2MASS source (J$08484864-4332578$) which is likely to be a reddened main sequence star and therefore may alternatively be a highly-variable radio star. On the basis of these results however, no definitive classification could be made.

With regards to lobe proper motions, only the morphologically complex sources (i.e. A1, A2, A3 and A4 of G310.1420+00.7583A and A2, B1 and B2 of G313.7654$-$00.8620) showed significant proper motions in the fitted positions for the lobes. Derived velocities for some of the lobes (i.e. $\sim10^4\kmps$ for B2 and A2 of G313.7654$-$00.8620) are extremely high and, considering typical, proper-motion values in the literature \citep[$300-1000\kmps$,][]{Marti1995,Rodriguez2008, Guzman2010}, likely due to errors during deconvolution and/or using the \textsc{imfit} task (as with flux variability). Towards G310.1420+00.7583A's lobe C (which is located away from differing $uv$ coverage effects) however, a more reasonable, and statistically-significant, proper motion of $1806\pm596\kmps$ is calculated. A derived position angle of $70\pm26\degr$ places the proper motion along the jet's propagation axis supporting the case that the observed motion is real. This is therefore the only radio lobe for which we reliably detect proper motions amongst the sample.

\subsection{Precession}
\label{sec:precession}
Various forms of evidence for precession are found towards all 4 objects of the sample with a large range in both derived precession periods and angles.

In the case of G263.7434$+$00.1161 the deconvolved position angles for the thermal jet (S) of $6\pm2\degr$, at both $6$ and $9\GHz$, suggest that the jet's current axis is offset by $35.7\pm1.6\degr$ w.r.t. the axis running through N, S and SE ($\theta_{PA}=-29.9\pm0.24\degr$). Assuming a jet velocity of $500\kmps$, $i = 90\degr$ and N to be optically thin emission from jet material (or a shock) which was therefore ejected on a ballistic trajectory from S $33\pm10\yr$ ago (assuming a distance error of $200\pc$), a precession rate of $1.1\degr\yr^{-1}$ is estimated, much higher than precession rates for other ionized jets found in the literature. The complex, $2.122\micron,\,{\rm H_2}$ emission previously observed \citep[Figure A9 of][]{deLuca2007} could be the result of a jet with a high precession rate creating shock sites over a wide range in angles, over a relatively short period, or multiple sources. It must be conceded that YSO/outflow multiplicity is an alternative explanation for these results. However, assuming that this change in the jet axis' angle over $33\yr$ is a result of a binary interaction and half the total period, a separation of $30\pm6\au$ is inferred for the companion. To calculate this orbital separation, we assume standard, Keplerian behaviour and employ the following equation:

\begin{align}
r&=\left(\frac{P^2 G M}{4\pi^2}\right)^\frac{1}{3},
\label{eq:orbitalradii}
\end{align}

\noindent where $P$ is period, $M$ is the total mass of the binary system and $r$ is the orbital semi-major axis.

For G310.0135+00.3892, variations in the position angles of the Herbig-Haro radio lobes HH1, HH2 and HH3, with respect to S, are seen. Fitting the $6\GHz$ peak positions of these HH objects, and the SW component with a jet model, via minimization of $\chi^2$ as described in appendix \ref{sec:appJetModel} \citep[with fixed values of $i=42\degr$ and $\theta_{PA}=31\degr$,][]{Boley2016}, yields a precession angle and period of $6^{+1}_{-2}\degr$ and $15480^{+3409}_{-2248}\yr$ respectively. Should this precession be due to a regular orbiting body, it should be separated by $1797^{+275}_{-191}\au$ from the MYSO. It must be noted that a precession angle and period of $8\degr$ and $8300\yr$ (inferring an orbital radius of $1200\au$) fit the (limited) data equally as well, but a period of $15480\yr$ represents the simplest model. Since the position angle for the established HH jet lies at an angle of $29\degr$, as discussed previously, the exact nature of the lobe N is an open question since it lies at a position angle of $-3\degr$ from the MYSO/thermal jet at S. If N is an optically thin \textsc{Hii} region, on the basis of its radio flux (and assuming a distance of $3.2\kpc$), a bolometric luminosity of $\sim4000\Lsol$ \citep[ZAMS type B2,][]{Davies2011} is implied, with a calculated average emission measure of $\sim7\times10^6\,\rm{pc\,cm}^{-6}$ and electron density of $\sim5000\cmc$, both of which are possibly too low for such a highly-compact ($\sim650\au$) \textsc{Hii} region. On the other hand, if a more rapid and wider angle precession is present compared to that found by analysis of the radio HH lobes, then this may support N as being the site of shock emission from the jet. However, SE lies along the accepted jet outflow axis and is roughly at the same separation from S as N is. This suggests that if a wider angle precession is present, it is only affecting the northern jet. It is interesting to note that \citet{Wheelwright2012} see this lobe asymmetry in their mid-IR observations too. Unfortunately 2MASS and GLIMPSE imagery is saturated, with image defects at N's position prohibiting its nature from being established any further.

\begin{figure}
\centering
\includegraphics[width=\columnwidth]{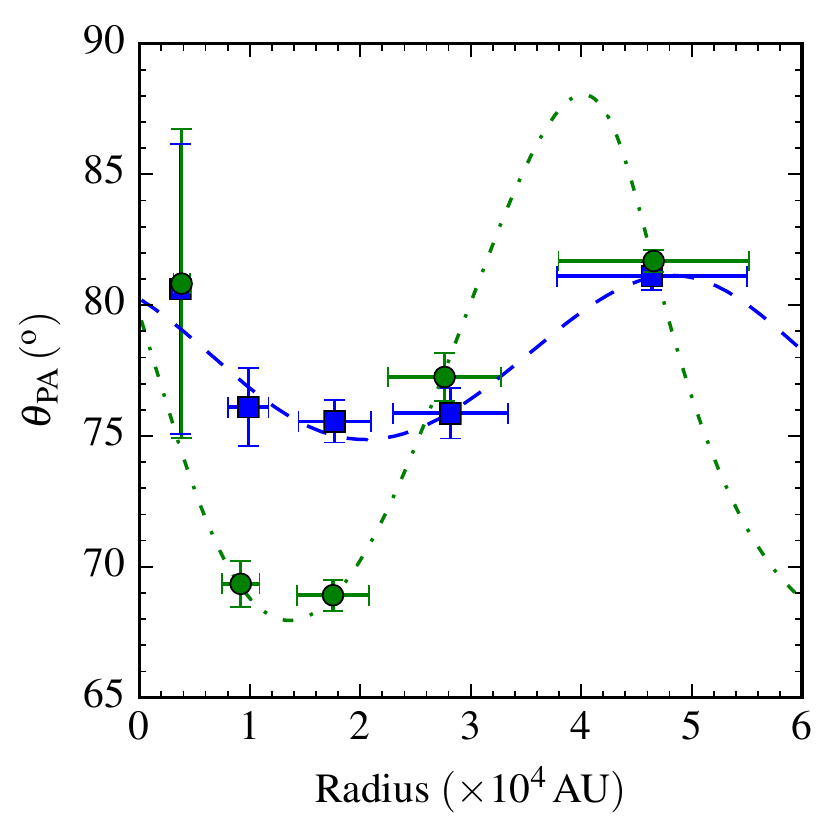} 
\caption{A plot of the radii and position angle of components A2, A3, A4, C and D of G310.1420+00.7583A, relative to A1, during both the 2013 (green circles) and 2014 (blue squares) epochs with their fitted curves (green dot-dashed line and blue dashed line respectively). Parameters derived for each curve are listed in \autoref{tab:jetmodelparams}.}
\label{fig:g310_RvsPA}
\end{figure}

\begin{figure}
\centering
\includegraphics[width=\columnwidth]{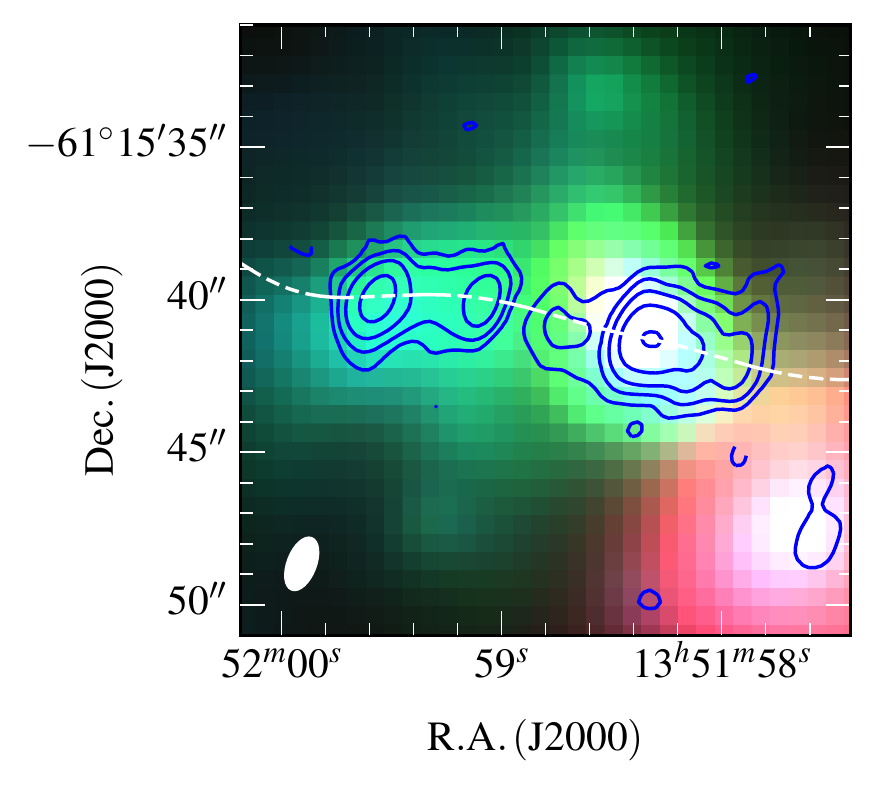}
\caption{A GLIMPSE RGB image of G310.1420+00.7583A overlayed with contours of the $9\GHz$ radio flux (contour levels are the same as in panel b.) of \autoref{fig:g310_1420contour}).The jet model uses the fitted values for the 2014 data (see \autoref{fig:g310_RvsPA}).}
\label{fig:g310_Jet_IR}
\end{figure}

\begin{table}
\caption{A table of the derived parameters from $\chi^2$ fitting of model in parameter space. Errors are calculated by consideration of the distribution of $\chi^2$ in parameter space. Any value with a $\,^\star$ next to it was held as a fixed parameter, so that the number of degrees of freedom, $\nu = N_\mathrm{points} - N_\mathrm{op}$ ($N_\mathrm{op}$ is the number of open parameters in the fit) is always $\> 1$.}
\begin{tabularx}{\columnwidth}{lYYcccY}
\hline 
\textbf{Source} & \textbf{Year} & \boldmath$\theta_{Pr.}$ & \boldmath$P_{Pr.}$ & \boldmath$\phi$ & \boldmath$i$ & \boldmath$\theta_{PA}$ \\
 & & $(\degr)$ & $(\yr)$ & $(\degr)$ & $(\degr)$ & $(\degr)$\\
\hline
\vspace{-2mm}&&&&&&\\
G310.0135 & 2014 & $6^{+1}_{-2}$ & $15480^{+3409}_{-2248}$ & $308^{+43}_{-26}$ & $42^\star$ & $31^\star$\vspace{2mm}\\
G310.1420 & 2013 & $19^{+2}_{-1}$ & $496^{+27}_{-34}$ & $340^{+15}_{-7}$ & $19^{+3}_{-10}$ & $78^\star$\vspace{2mm}\\
 & 2014 & $6^{+3}_{-3}$ & $568^{+318}_{-126}$ & $313^{+88}_{-57}$ & $16^{+26}_{-16}$ & $78^\star$\vspace{2mm}\\
G313.7654 & 2014 & $12^{+2}_{-1}$ & $920^{+82}_{-32}$ & $338^{+11}_{-26}$ & $51^{+3}_{-2}$ & $121^\star$\\
\hline 
\end{tabularx} 
\label{tab:jetmodelparams}
\end{table}

In \autoref{fig:g310_RvsPA} the position angle of the peak emission for each of G310.1420+00.7583A's lobes is plotted as a function of radius and position angle from A1 (i.e. the MYSO), which shows evidence for precession in the jet's axis. These positions were fitted with a precessing jet model with inclination left as a free parameter. For both the 2013 and 2014 data, the best fitting model's parameters are tabulated in \autoref{tab:jetmodelparams}. In \autoref{fig:g310_Jet_IR} this model derived for the 2014 data is plotted over GLIMPSE, MIR images. It can be seen that the jet model's path traces both the east, MIR $4.5\micron$ excess (EGO) and general, extended, radio morphology well. Comparison of the 2013 and 2014 best fit models shows the derived quantities largely agree, apart from the precession angles ($19\degr$ and $6\degr$ for the 2013/2014 data respectively). This discrepancy between the two models is likely due to deconvolution/imaging errors discussed in \autoref{sec:atcafluxrecovery} and \autoref{sec:var+pms}. As with G310.0135+00.3892, assuming that the jet precession is due to a binary whose orbital period is the same as the precession period, an orbital radius of $141^{+7}_{-8}\au$ or $155^{+58}_{-24}\au$ is inferred, depending on whether the fit parameters for the 2013 or 2014 data are used respectively.

\begin{figure}
\centering
\includegraphics[width=\columnwidth]{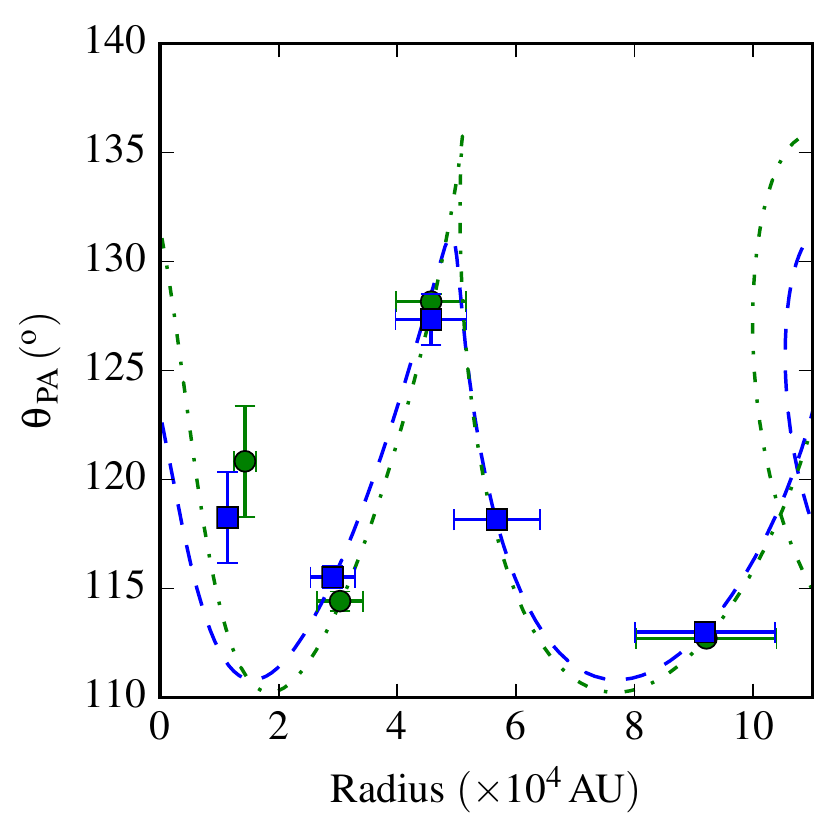} 
\caption{A plot of the radii and position angles for components A2, B1, B3, C and D of G313.7654--00.8620 (relative to A1) where symbols and fitted lines have the same meaning as in \autoref{fig:g310_Jet_IR}. For 2013 and 2014 data, values for $\theta_{PA}$ of $123\degr$ and $121\degr$, $P_{pr}$ of $950$ and $920\yr$, $i$ of $54\degr$ and $51\degr$, $\theta_{pr}$ of $15\degr$ and $12\degr$ and $\phi$ of $35\degr$ and $338\degr$ were used for the 2013 and 2014 models respectively.}
\label{fig:g313_RvsPA}
\end{figure}

\begin{figure}
\centering
\includegraphics[width=\columnwidth]{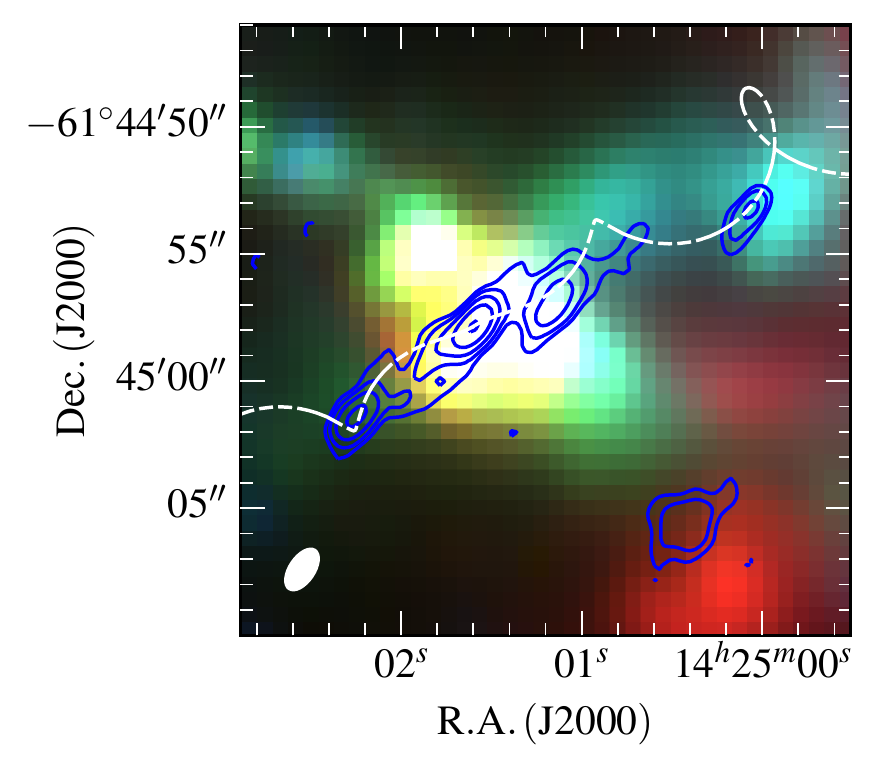} 
\caption{A GLIMPSE RGB image of G313.7654--00.8620 overlayed with contours of the $9\GHz$ radio flux (contour levels of $-4, 4, 8, 15, 28$ and $54\sigma$ where $\sigma = 9.0\uJy\,\mathrm{beam}^{-1}$). The jet model uses the fitted values for the 2014 data (see \autoref{fig:g313_RvsPA}).}
\label{fig:g313_Jet_IR}
\end{figure}

In the case of G313.7654$-$00.8620, the radii and position angles of A2, B1, B3, C and D were calculated with respect to the MYSO, A1 (plotted in \autoref{fig:g313_RvsPA}). Because only the 2014 images detected B3, fitting with inclination as a free parameter was only attempted towards the 2014 lobe positions (in order to have at least one degree of freedom). Component B2 is neglected from this analysis due to its unknown nature and its deviation from the position angles of the other lobes. Fitting yields a precessing jet model with a precession angle of $12^{+2}_{-1}\degr$, period of $920^{+82}_{-32}\yr$ and inclination of $51^{+3}_{-2}\degr$ which is shown overlayed upon a GLIMPSE RGB image in \autoref{fig:g313_Jet_IR}. As with the other objects, if a binary companion has induced this precession, a separation for the binary companion of $270^{+19}_{13}\au$ is inferred. The fitted model does not accurately trace the MIR $4.5\micron$ excess, in comparison with G310.1420+00.7583A, the diffuse radio emission detected between A2 and C, or the radio lobe, G \citep[which is coincident with the H$_2$ $2.122\micron$ emission lobe, `knot 4' from][]{Caratti2015}. It is possible that G313.7654$-$00.8620 is a more complex system, and (in addition to periodic precession) a large overall shift in the precession axis may have occurred, on account of G's position and the results of \citet{Caratti2015}, from $\sim98\degr$ to $121\degr$ over a period of approximately $4100\yr$, or $5.6\times10^{-3}\,\degr\yr^{-1}$. 

\begin{figure}
\includegraphics[width=\columnwidth]{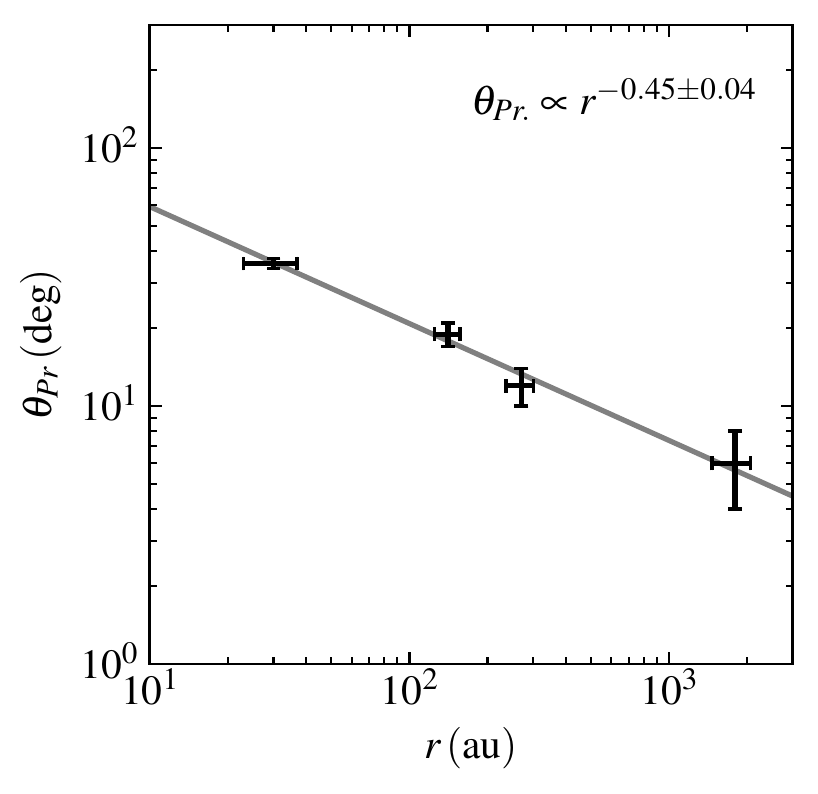}
\caption{A plot of the precession angle ($\theta_\mathrm{Pr}$) against inferred orbital radii ($r$). Orthogonal distance regression (taking into account errors in both variables) was used to fit the power law which is plotted in grey and explicitly stated in \autoref{eq3:prangorbradplaw}.}
\label{fig:PrAngVsOrbRadii}
\end{figure}

In \autoref{fig:PrAngVsOrbRadii} the four, inferred precession angles for the sample, and the relevant orbital radii, are plotted. Fitted power laws for precession angle against both period and inferred orbital radius are explicitly stated in \autoref{eq3:prangprperiodplaw} and \autoref{eq3:prangorbradplaw} respectively, below:

\begin{align}
\log_{10}\left(\theta_\mathrm{Pr}\right)&=\left(2.20\pm0.13\right)-\left(0.36\pm0.05\right)\log_{10}\left(\mathrm{P_{Pr}}\right)\label{eq3:prangprperiodplaw}\\
\log_{10}\left(\theta_\mathrm{Pr}\right)&=\left(2.23\pm0.08\right)-\left(0.45\pm0.04\right)\log_{10}\left(r\right)\label{eq3:prangorbradplaw}
\end{align}

\noindent These relations show that for shorter precession periods/smaller orbital radii for the hypothesised binary companions, a larger precession angle is observed. This agrees with the idea that a closer companion would deflect a jet's stream to a greater degree, either through gravitational influence upon the ballistic trajectory of jet material, or alteration of the disc's magnetic field. However to establish this explanation would require both a larger sample than that presented here and direct confirmation of the existence of binary companions.

\section{Summary and Conclusions}
\label{sec:summary}
Sensitive radio observations at $6$ and $9\GHz$ were conducted towards a sample of 4 MYSOs known to harbour ionized jets associated with shock-ionized lobes. This sample was selected from the previous work of \citetalias{Purser2016} in order to investigate their temporal evolution, faint emission and precession of their outflow axes. From careful consideration of the observational differences between each epoch's dataset, their subsequent comparison as well as modelling of the relative lobe positions from their sourcing jets, the following can be concluded:

\begin{itemize}
\item Fainter emission is detected towards 3 of 4 jets, which is sourced directly from the thermal jet itself in one object (G263.7434+00.1161), radio Herbig-Haro objects in G310.0135+00.3892 and extra non-thermal emission in G313.7654$-$00.8620. This suggests that mass loss in MYSO jets is always occurring, though at highly variable rates.
\item Proper motion is detected at the $>3\sigma$ level towards one lobe (C) of G310.1420+00.7583A, with a derived velocity of $1806\pm596\kmps$ parallel to the jet's propagation axis.
\item Change in the flux morphology is seen towards a non-thermal lobe (D) of G310.1420+00.7583A, supporting the hypothesis that the emission is the result of an evolving shock, rather than direct emission from the jet's stream.
\item Evidence for precession is found in all objects within the sample, with inferred precession angles ranging from $6\pm3\degr$ to $36\pm2\degr$ and periods from $66\pm20$ to $15480^{+3409}_{-2248}\yr$. If the assumed precession of jet outflow axes is caused by the bound orbits of binary companions, estimated orbital radii of $30\pm6\au$, $1797^{+275}_{-191}\au$, $141^{+7}_{-8}\au$ and $270^{+19}_{-13}\au$ are found for G263.7434+00.1161, G310.0135+00.3892, G310.1420+00.7583A and G313.7654$-$00.8620 respectively.
\end{itemize}

Considering the apparent precessions of the MYSOs observed, it is clear that these 4 ionized jets are not restricted to a single outflow angle over time. Compared to precession seen towards low-mass examples, it is both more extreme and also more rapid. Definitive variability or proper motions were not observed over a 2 year period towards the shock ionized lobes or thermal jets, however future studies, with longer time baselines, towards a larger sample of objects from \citetalias{Purser2016} will inform more reliably on these matters. The results presented here therefore form the foundation for such studies and hint that it will produce interesting results in the future.

\section*{Acknowledgements}
SJDP gratefully acknowledges the studentship funded by the Science and Technology Facilities Council of the United Kingdom (STFC). This paper has made use of information from the RMS survey database at http://rms.leeds.ac.uk/ which was constructed with support from the Science and Technology Facilities Council of the United Kingdom.

\bibliographystyle{mn2e}
\bibliography{Biblio}

\onecolumn
\appendix
\begin{landscape}
\section{Supplementary Tables}
\label{app:tables}
\begin{table}
\centering
\caption{A table of lobe positions, integrated fluxes and deconvolved dimensions at $9\GHz$ for the 2013 epoch. The final column indicates which robustness was employed in the clean map.}
\begin{tabular}{lcccccccccc}
\hline
\textbf{Source} & \textbf{Lobe} & \boldmath$\alpha$ & \boldmath$\delta\alpha$ & \boldmath$\delta$ & \boldmath$\delta\delta$ & \boldmath$S_9$ & \boldmath$\theta_{Maj}$ & \boldmath$\theta_{Min}$ & \boldmath$\theta_{PA}$ & \textbf{R} \\
 & & (J2000) & $(\arcsec)$ & (J2000) & $(\arcsec)$ & $(\uJy)$ & $(\arcsec)$ & $(\arcsec)$ & $ (\degr)$  & \\ 
 \hline
G263.7434+00.1611 & N & $08\rahr48\ramin48.47\rasec$ & $0.019$ & $-43\degr32\arcmin24.0\arcsec$ & $0.064$ & $615\pm79$ & $1.32\pm0.31$ & $0.55\pm0.14$ & $175.0\pm13.0$ & 0.5 \\
 & S2 & - & - & - & - & $<80\pm27$ & $-$ & $-$ & $-$ & 0.5 \\
 & S & $08\rahr48\ramin48.66\rasec$ & $0.014$ & $-43\degr32\arcmin28.7\arcsec$ & $0.068$ & $1026\pm110$ & $2.34\pm0.22$ & $0.59\pm0.12$ & $6.1\pm3.2$ & 0.5 \\
 & SE & $08\rahr48\ramin48.87\rasec$ & $0.206$ & $-43\degr32\arcmin32.9\arcsec$ & $0.596$ & $206\pm87$ & $3.41\pm1.91$ & $1.03\pm0.94$ & $163.0\pm41.0$ & 2 \\
G310.0135+00.3892 & N & $13\rahr51\ramin37.85\rasec$ & $0.013$ & $-61\degr39\arcmin06.3\arcsec$ & $0.012$ & $709\pm85$ & $<0.35$ & $<0.22$ & $-$ & $-1$ \\
 & S & $13\rahr51\ramin37.85\rasec$ & $0.032$ & $-61\degr39\arcmin07.8\arcsec$ & $0.022$ & $413\pm76$ & $0.41\pm0.16$ & $0.16\pm0.15$ & $79.0\pm83.0$ & $-1$ \\
 & SW & $13\rahr51\ramin37.73\rasec$ & $0.055$ & $-61\degr39\arcmin09.1\arcsec$ & $0.119$ & $149\pm65$ & $-$ & $-$ & $-$ & $-1$ \\
G310.1420+00.7583A & A1 & $13\rahr51\ramin58.37\rasec$ & $0.008$ & $-61\degr15\arcmin41.2\arcsec$ & $0.007$ & $1848\pm152$ & $0.55\pm0.04$ & $0.45\pm0.04$ & $92.0\pm21.0$ & $-1$ \\
 & A2 & $13\rahr51\ramin58.26\rasec$ & $0.006$ & $-61\degr15\arcmin41.5\arcsec$ & $0.008$ & $909\pm83$ & $-$ & $-$ & $-$ & $-1$ \\
 & A3 & $13\rahr51\ramin58.16\rasec$ & $0.021$ & $-61\degr15\arcmin41.8\arcsec$ & $0.015$ & $1938\pm182$ & $1.27\pm0.06$ & $0.46\pm0.05$ & $58.4\pm2.1$ & $-1$ \\
 & A4 & $13\rahr51\ramin57.94\rasec$ & $0.018$ & $-61\degr15\arcmin42.3\arcsec$ & $0.035$ & $660\pm97$ & $0.85\pm0.12$ & $0.28\pm0.17$ & $1.9\pm7.0$ & $-1$ \\
 & B & $13\rahr51\ramin58.71\rasec$ & $0.068$ & $-61\degr15\arcmin40.7\arcsec$ & $0.052$ & $311\pm66$ & $-$ & $-$ & $-$ & $0.5$ \\
 & C & $13\rahr51\ramin59.10\rasec$ & $0.044$ & $-61\degr15\arcmin40.0\arcsec$ & $0.024$ & $1306\pm139$ & $2.00\pm0.13$ & $0.53\pm0.17$ & $78.3\pm3.0$ & $0.5$ \\
 & D & $13\rahr51\ramin59.55\rasec$ & $0.018$ & $-61\degr15\arcmin40.0\arcsec$ & $0.017$ & $2139\pm178$ & $1.44\pm0.07$ & $1.00\pm0.09$ & $97.0\pm12.0$ & $0.5$ \\

G313.7654$-$00.8620 & A1 & $14\rahr25\ramin01.59\rasec$ & $0.013$ & $-61\degr44\arcmin57.8\arcsec$ & $0.018$ & $603\pm58$ & $0.43\pm0.23$ & $0.25\pm0.14$ & $47.0\pm64.0$ & 0.5 \\
 & A2 & $14\rahr25\ramin01.81\rasec$ & $0.065$ & $-61\degr44\arcmin58.8\arcsec$ & $0.150$ & $165\pm46$ & $1.48\pm0.53$ & $0.24\pm0.50$ & $164.0\pm26.0$ & $0.5$ \\
 & B1 & $14\rahr25\ramin01.10\rasec$ & $0.086$ & $-61\degr44\arcmin55.9\arcsec$ & $0.085$ & $299\pm62$ & $1.58\pm0.36$ & $0.85\pm0.48$ & $125.0\pm23.0$ & 0.5 \\
 & B2 & $14\rahr25\ramin01.21\rasec$ & $0.059$ & $-61\degr44\arcmin57.9\arcsec$ & $0.082$ & $145\pm36$ & $0.58\pm0.48$ & $0.36\pm0.26$ & $40.0\pm47.0$ & 0.5 \\
 & B3 & $14\rahr25\ramin00.72\rasec$ & $0.122$ & $-61\degr44\arcmin55.1\arcsec$ & $0.192$ & $88\pm38$ & $0.96\pm0.74$ & $0.55\pm0.36$ & $18.0\pm35.0$ & 0.5 \\
 & C & $14\rahr25\ramin02.25\rasec$ & $0.031$ & $-61\degr45\arcmin01.3\arcsec$ & $0.038$ & $444\pm58$ & $0.85\pm0.22$ & $0.82\pm0.28$ & $72.0\pm86.0$ & 0.5 \\
 & D & $14\rahr25\ramin00.07\rasec$ & $0.043$ & $-61\degr44\arcmin53.3\arcsec$ & $0.057$ & $160\pm33$ & $<0.72$ & $<0.29$ & $-$ & 0.5 \\
\hline
\end{tabular}
\label{tab:JetsLobesPositions9GHz2013}
\end{table}
\end{landscape}

\small
\begin{landscape}
\begin{table}
\centering
\caption{A table of lobe positions, integrated fluxes and deconvolved dimensions at 6 GHz for the 2014 epoch. The final column indicates which robustness was employed in the clean map. Those components with a $^1$ above their name are extended and therefore their peak flux positions are given, whose errors cover half of their $3\sigma$ spatial extent. Their provided fluxes are integrated over the $3\sigma$ contours. }
\begin{tabular}{lcccccccccc}
\hline
\textbf{Source} & \textbf{Lobe} & \boldmath$\alpha$ & \boldmath$\delta\alpha$ & \boldmath$\delta$ & \boldmath$\delta\delta$ & \boldmath$S_6$ & \boldmath$\theta_{Maj}$ & \boldmath$\theta_{Min}$ & \boldmath$\theta_{PA}$ & \textbf{R} \\
 & & (J2000) & $(\arcsec)$ & (J2000) & $(\arcsec)$ & $(\uJy)$ & $(\arcsec)$ & $(\arcsec)$ & $ (\degr)$  & \\ 
 \hline
G263.7434+00.1161 & N & $08\rahr48\ramin48.48\rasec$ & $0.005$ & $-43\degr32\arcmin23.9\arcsec$ & $0.027$ & $626\pm43$ & $1.40\pm0.19$ & $0.50\pm0.03$ & $177.1\pm3.6$ & $0.5$\\
 & S2 & $08\rahr48\ramin48.63\rasec$ & $0.009$ & $-43\degr32\arcmin57.7\arcsec$ & $0.050$ & $319\pm28$ & $1.13\pm0.49$ & $0.35\pm0.17$ & $168.0\pm21.0$ & $0.5$ \\
 & S & $08\rahr48\ramin48.65\rasec$ & $0.004$ & $-43\degr32\arcmin28.5\arcsec$ & $0.029$ & $767\pm52$ & $2.24\pm0.13$ & $0.62\pm0.05$ & $5.9\pm2.0$ &$ 0.5$ \\
 & SE & $08\rahr48\ramin48.83\rasec$ & $0.035$ & $-43\degr32\arcmin33.4\arcsec$ & $0.365$ & $126\pm22$ & $5.18\pm1.09$ & $0.64\pm0.23$ & $178.7\pm2.7$ & $0.5$ \\
G310.0135+00.3892 & N & $13\rahr51\ramin37.87\rasec$ & $0.005$ & $-61\degr39\arcmin06.2\arcsec$ & $0.009$ & $707\pm53$ & $0.28\pm0.07$ & $0.15\pm0.12$ & $87.0\pm75.0$ & -1 \\
 & S & $13\rahr51\ramin37.87\rasec$ & $0.013$ & $-61\degr39\arcmin07.7\arcsec$ & $0.022$ & $347\pm37$ & $0.50\pm0.16$ & $0.35\pm0.20$ & $155.0\pm89.0$ & -1 \\
 & SW & $13\rahr51\ramin37.73\rasec$ & $0.135$ & $-61\degr39\arcmin09.3\arcsec$ & $0.168$ & $126\pm35$ & $<2.60$ & $<0.51$ & $-$ & -1 \\
 & HH0\footnotemark[1] & $13\rahr51\ramin41.84\rasec$ & $2.4$ & $-61\degr38\arcmin18.7\arcsec$ & $3.3$ & $145\pm19$ & $-$ & $-$ & $-$ & 0.5 \\
 & HH1\footnotemark[1] & $13\rahr51\ramin48.13\rasec$ & $2.7$ & $-61\degr36\arcmin46.2\arcsec$ & $3.2$ & $422\pm48$ & $-$ & $-$ & $-$ & 0.5 \\
 & HH2\footnotemark[1] & $13\rahr51\ramin55.35\rasec$ & $1.6$ & $-61\degr35\arcmin38.6\arcsec$ & $1.8$ & $121\pm21$ & $-$ & $-$ & $-$ & 0.5 \\
G310.1420+00.7583A & A1+A2 & $13\rahr51\ramin58.33\rasec$ & $0.004$ & $-61\degr15\arcmin41.2\arcsec$ & $0.006$ & $2744\pm177$ & $0.84\pm0.03$ & $0.55\pm0.05$ & $62.6\pm5.4$ & -1 \\
 & A3 & $13\rahr51\ramin58.18\rasec$ & $0.006$ & $-61\degr15\arcmin41.7\arcsec$ & $0.010$ & $2053\pm144$ & $1.09\pm0.04$ & $0.42\pm0.06$ & $37.5\pm3.1$ & -1 \\
 & A4 & $13\rahr51\ramin57.96\rasec$ & $0.011$ & $-61\degr15\arcmin42.0\arcsec$ & $0.021$ & $888\pm83$ & $0.93\pm0.10$ & $0.43\pm0.15$ & $37.8\pm9.6$ & -1 \\
 & B & $13\rahr51\ramin58.70\rasec$ & $0.027$ & $-61\degr15\arcmin41.0\arcsec$ & $0.102$ & $369\pm51$ & $2.24\pm0.36$ & $0.71\pm0.21$ & $175.8\pm6.6$ & 0.5 \\
 & C & $13\rahr51\ramin59.12\rasec$ & $0.018$ & $-61\degr15\arcmin40.0\arcsec$ & $0.020$ & $1508\pm115$ & $1.91\pm0.09$ & $1.09\pm0.18$ & $80.7\pm5.8$ & 0.5 \\
 & D & $13\rahr51\ramin59.55\rasec$ & $0.006$ & $-61\degr15\arcmin40.1\arcsec$ & $0.011$ & $2474\pm157$ & $1.31\pm0.05$ & $1.15\pm0.07$ & $136.0\pm16.0$ & 0.5 \\
G313.7654$-$00.8620 & A1 & $14\rahr25\ramin01.59\rasec$ & $0.012$ & $-61\degr44\arcmin57.9\arcsec$ & $0.024$ & $338\pm30$ & $-$ & $-$ & $-$ & 0.5 \\
 & A2 & $14\rahr25\ramin01.69\rasec$ & $0.076$ & $-61\degr44\arcmin58.2\arcsec$ & $0.064$ & $426\pm47$ & $3.42\pm0.28$ & $0.70\pm0.28$ & $122.6\pm3.1$ & 0.5 \\
 & B1 & $14\rahr25\ramin01.10\rasec$ & $0.044$ & $-61\degr44\arcmin56.0\arcsec$ & $0.043$ & $231\pm28$ & $-$ & $-$ & $-$ & 0.5 \\
 & B2 & $14\rahr25\ramin01.18\rasec$ & $0.049$ & $-61\degr44\arcmin57.5\arcsec$ & $0.047$ & $213\pm27$ & $-$ & $-$ & $-$ & 0.5 \\
 & B3 & $14\rahr25\ramin00.71\rasec$ & $0.160$ & $-61\degr44\arcmin54.6\arcsec$ & $0.176$ & $118\pm30$ & $2.15\pm0.90$ & $1.20\pm0.77$ & $122.0\pm51.0$ & 0.5 \\
 & D & $14\rahr25\ramin00.09\rasec$ & $0.043$ & $-61\degr44\arcmin53.3\arcsec$ & $0.060$ & $165\pm24$ & $<1.20$ & $<0.84$ & $-$ & 0.5 \\
 & F & $14\rahr25\ramin04.70\rasec$ & $0.064$ & $-61\degr44\arcmin50.3\arcsec$ & $0.185$ & $85\pm22$ & $<2.2$ & $<0.77$ & $-$ & 0.5 \\
 & G & $14\rahr24\ramin57.07\rasec$ & $0.581$ & $-61\degr44\arcmin53.2\arcsec$ & $0.236$ & $171\pm59$ & $4.75\pm1.56$ & $1.09\pm0.88$ & $101\pm15$ & 0.5 \\ 
\hline
\end{tabular}
\label{tab:JetsLobesPositions6GHz}
\end{table}
\end{landscape}

\small
\begin{landscape}
\begin{table}
\centering
\caption{A table of lobe positions, integrated fluxes and deconvolved dimensions at 9 GHz for the 2014 epoch. The final column indicates which robustness was employed in the clean map.}
\begin{tabular}{lcccccccccc}
\hline
\textbf{Source} & \textbf{Lobe} & \boldmath$\alpha$ & \boldmath$\delta\alpha$ & \boldmath$\delta$ & \boldmath$\delta\delta$ & \boldmath$S_9$ & \boldmath$\theta_{Maj}$ & \boldmath$\theta_{Min}$ & \boldmath$\theta_{PA}$ & \textbf{R} \\
 & & (J2000) & $(\arcsec)$ & (J2000) & $(\arcsec)$ & $(\uJy)$ & $(\arcsec)$ & $(\arcsec)$ & $ (\degr)$  & \\ 
\hline
G263.7434+00.1161 & N & $08\rahr48\ramin48.48\rasec$ & $0.004$ & $-43\degr32\arcmin23.8\arcsec$ & $0.026$ & $594\pm44$ & $1.40\pm0.14$ & $0.39\pm0.03$ & $176.1\pm2.0$ & $0.5$ \\
 & Radio Star & $08\rahr48\ramin48.63\rasec$ & $0.007$ & $-43\degr32\arcmin57.7\arcsec$ & $0.037$ & $363\pm31$ & $1.04\pm0.27$ & $0.27\pm0.07$ & $167.8\pm7.8$ & $0.5$ \\
 & S & $08\rahr48\ramin48.66\rasec$ & $0.003$ & $-43\degr32\arcmin28.5\arcsec$ & $0.023$ & $837\pm58$ & $1.85\pm0.10$ & $0.43\pm0.04$ & $5.8\pm1.6$ & $0.5$ \\
 & SE & $08\rahr48\ramin48.82\rasec$ & $0.047$ & $-43\degr32\arcmin33.1\arcsec$ & $0.468$ & $135\pm29$ & $5.29\pm1.27$ & $0.73\pm0.18$ & $177.2\pm2.4$ & $0.5$ \\
G310.0135+00.3892 & N & $13\rahr51\ramin37.86\rasec$ & $0.005$ & $-61\degr39\arcmin06.2\arcsec$ & $0.013$ & $695\pm66$ & $<0.46$ & $<0.06$ & $-$ & $-1$ \\
 & S & $13\rahr51\ramin37.87\rasec$ & $0.006$ & $-61\degr39\arcmin07.6\arcsec$ & $0.018$ & $481\pm53$ & $-$ & $-$ & $-$ & $-1$ \\
 & SW & $13\rahr51\ramin37.75\rasec$ & $0.045$ & $-61\degr39\arcmin08.9\arcsec$ & $0.078$ & $93\pm32$ & $-$ & $-$ & $-$ & $-1$ \\
G310.1420+00.7583A & A1 & $13\rahr51\ramin58.39\rasec$ & $0.006$ & $-61\degr15\arcmin41.2\arcsec$ & $0.014$ & $1462\pm125$ & $0.66\pm0.07$ & $0.28\pm0.03$ & $162.4\pm4.3$ & $-1$ \\
 & A2 & $13\rahr51\ramin58.28\rasec$ & $0.006$ & $-61\degr15\arcmin41.4\arcsec$ & $0.012$ & $1842\pm149$ & $0.61\pm0.07$ & $0.41\pm0.03$ & $171.0\pm12.0$ & $-1$ \\
 & A3 & $13\rahr51\ramin58.14\rasec$ & $0.024$ & $-61\degr15\arcmin41.7\arcsec$ & $0.045$ & $1295\pm160$ & $1.36\pm0.14$ & $0.81\pm0.07$ & $163.4\pm8.0$ & $-1$ \\
 & A4 & $13\rahr51\ramin57.93\rasec$ & $0.018$ & $-61\degr15\arcmin41.9\arcsec$ & $0.058$ & $570\pm94$ & $1.00\pm0.22$ & $0.32\pm0.15$ & $7.1\pm13.9$ & $-1$ \\
 & B & $13\rahr51\ramin58.72\rasec$ & $0.055$ & $-61\degr15\arcmin41.0\arcsec$ & $0.100$ & $357\pm69$ & $1.40\pm0.39$ & $0.75\pm0.64$ & $35.0\pm36.0$ & $0.5$ \\
 & C & $13\rahr51\ramin59.13\rasec$ & $0.037$ & $-61\degr15\arcmin40.0\arcsec$ & $0.031$ & $1154\pm120$ & $1.82\pm0.15$ & $0.82\pm0.33$ & $88.6\pm8.8$ & $0.5$ \\
 & D & $13\rahr51\ramin59.56\rasec$ & $0.001$ & $-61\degr15\arcmin40.0\arcsec$ & $0.013$ & $2079\pm151$ & $1.13\pm0.06$ & $0.74\pm0.11$ & $95.8\pm8.0$ & $0.5$ \\
G313.7654$-$00.8620 & A1 & $14\rahr25\ramin01.59\rasec$ & $0.012$ & $-61\degr44\arcmin57.9\arcsec$ & $0.015$ & $451\pm39$ & $-$ & $-$ & $-$ & 0.5 \\
 & A2 & $14\rahr25\ramin01.74\rasec$ & $0.165$ & $-61\degr44\arcmin58.4\arcsec$ & $0.150$ & $469\pm67$ & $4.90\pm0.56$ & $1.09\pm0.15$ & $130.6\pm2.1$ & 0.5 \\
 & B1 & $14\rahr25\ramin01.10\rasec$ & $0.061$ & $-61\degr44\arcmin56.2\arcsec$ & $0.071$ & $277\pm42$ & $1.62\pm0.34$ & $0.99\pm0.21$ & $132.0\pm25.0$ & 0.5 \\
 & B2 & $14\rahr25\ramin01.18\rasec$ & $0.068$ & $-61\degr44\arcmin57.5\arcsec$ & $0.081$ & $146\pm26$ & $<1.70$ & $<0.29$ & $-$ & 0.5 \\
 & B3 & $14\rahr25\ramin00.72\rasec$ & $0.195$ & $-61\degr44\arcmin54.8\arcsec$ & $0.218$ & $81\pm27$ & $1.97\pm1.11$ & $0.32\pm0.34$ & $134.0\pm33.0$ & 0.5 \\
 & C & $14\rahr25\ramin02.25\rasec$ & $0.023$ & $-61\degr45\arcmin01.4\arcsec$ & $0.028$ & $372\pm38$ & $0.87\pm0.21$ & $0.35\pm0.11$ & $125.0\pm19.0$ & 0.5 \\
 & D & $14\rahr25\ramin00.08\rasec$ & $0.038$ & $-61\degr44\arcmin53.4\arcsec$ & $0.056$ & $164\pm24$ & $-$ & $-$ & $-$ & 0.5 \\
 & F & $14\rahr25\ramin04.68\rasec$ & $0.104$ & $-61\degr44\arcmin50.1\arcsec$ & $0.105$ & $85\pm25$ & $-$ & $-$ & $-$ & 0.5 \\
 & G & $14\rahr24\ramin57.04\rasec$ & $0.130$ & $-61\degr44\arcmin54.1\arcsec$ & $0.125$ & $51\pm18$ & $-$ & $-$ & $-$ & 0.5 \\ 
\hline
\end{tabular}
\label{tab:JetsLobesPositions9GHz}
\end{table}
\end{landscape}

\clearpage
\section{Supplementary Figures}
\label{app:figures}
\begin{figure}
\begin{center}
\includegraphics[]{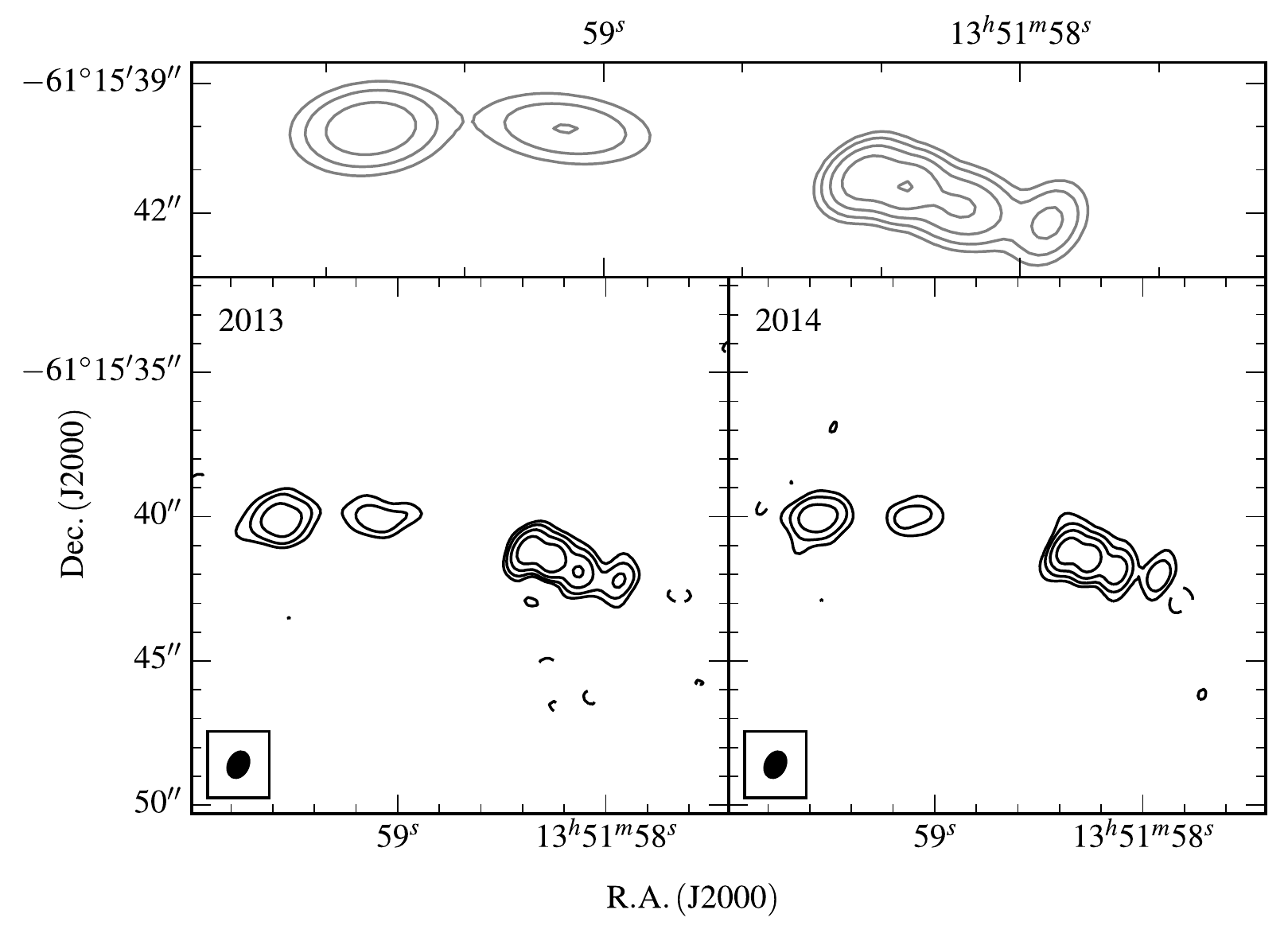}
\caption{Clean maps of the synthetic data at 9 GHz (bottom left and right plots). The top plot represents the model used to generate the synthetic visibilities convolved with the common beam used in both synthetic clean maps ($0.94\arcsec\times0.68\arcsec, \theta_{PA}=-26.3\degr$, shown bottom left of each clean map). All contours are logarithmically spaced from $3\sigma$ (where $\sigma = 23.0\uJy/\mathrm{beam}$, the noise in the 2013 synthetic clean map) to $95\%$ the maximum flux in the model's image, specifically $-3, 3, 6, 14, 29$ and $63\sigma$.}
\label{fig:SOcleanmaps}
\end{center}
\end{figure}

\begin{figure}
\begin{center}
\includegraphics[]{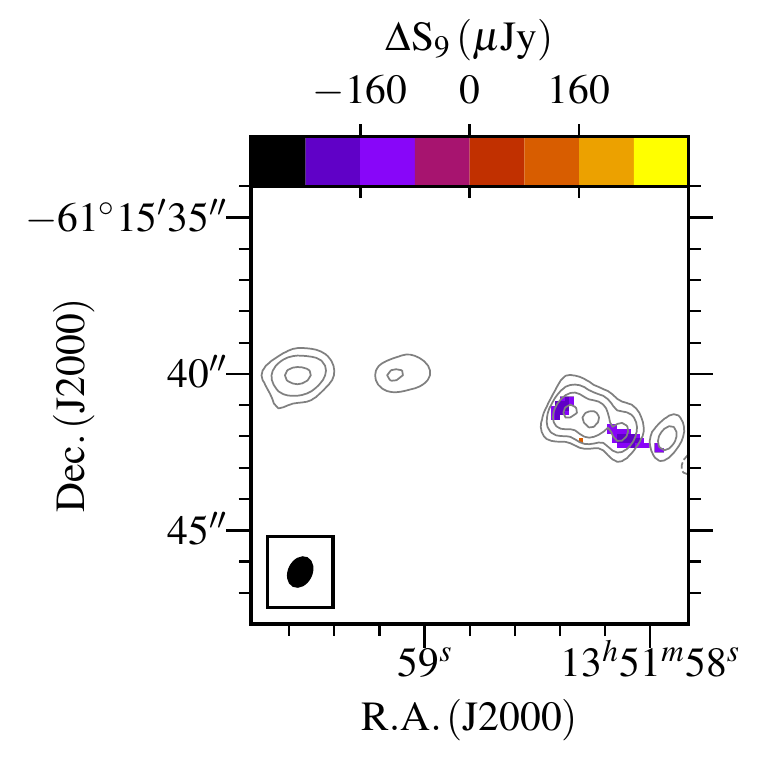} 
\caption{A map of pixel-to-pixel flux differences between 2013 and 2014 synthetic clean maps (colour-scale). The grey contours are the same as per \autoref{fig:g310_1420fd}.}
\label{fig:SOfluxdiff}
\end{center}
\end{figure}
\begin{figure}
\begin{center}
\includegraphics[]{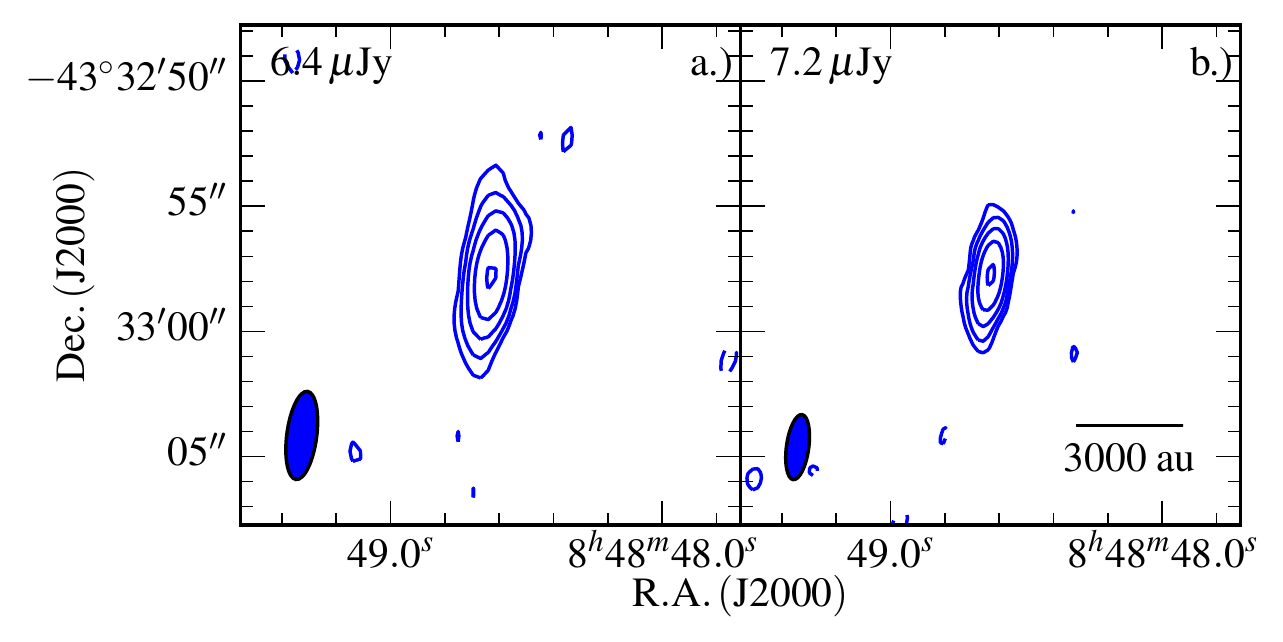} 
\caption{Primary beam corrected, radio contour maps of lobe S2 detected in the primary beam of our observation of G263.7434+00.1161 utilising a robustness of $0.5$ at: a.) $6\GHz$; b.) $9\GHz$. Restoring beams used are illustrated in the bottom left corners (same dimensions as in \autoref{fig:g263_7434contour}) and contours are $(-3, 3, 6, 11, 22, 44)\times\sigma$ and $(-3, 3, 6, 11, 22, 42)\times\sigma$ for the $6$ and $9\GHz$ data respectively, where $\sigma$ is the noise in the image indicated in the top-left corner of each sub-plot.}
\label{fig:g263_7434_Radio_Star}
\end{center}
\end{figure}
\begin{figure}
\begin{center}
\includegraphics[]{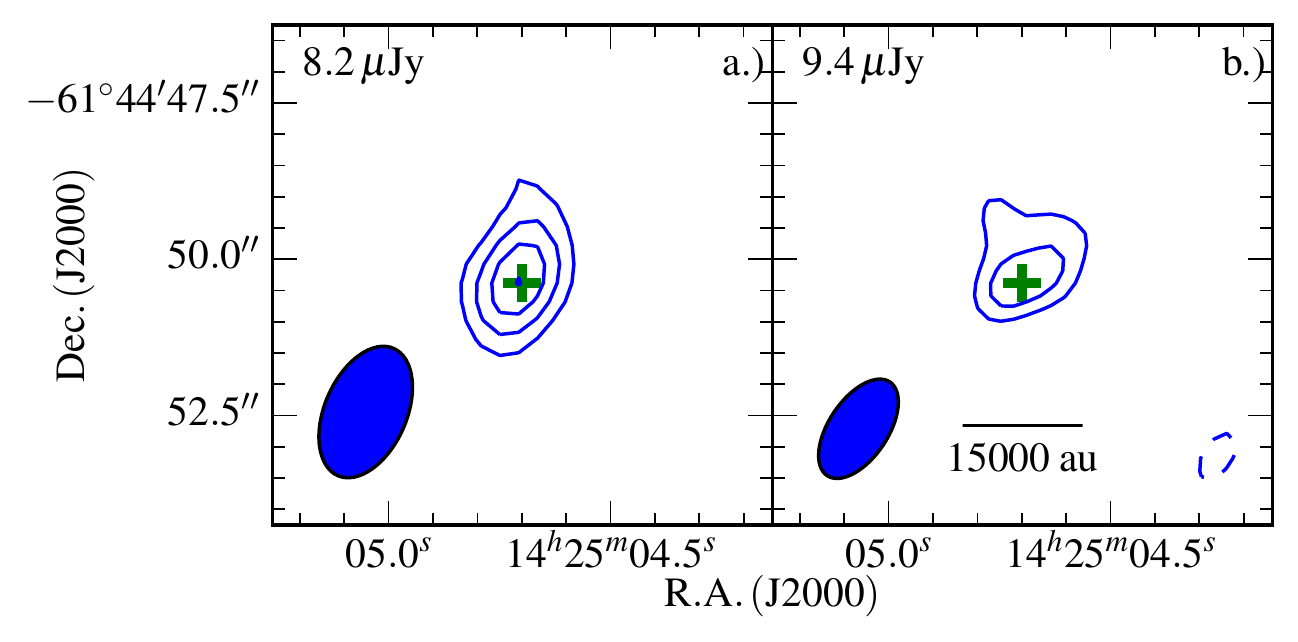} 
\caption{Primary beam corrected, radio contour maps of `F' towards G313.7654-00.8620 utilising a robustness of $0.5$ at: a.) $6\GHz$; b.) $9\GHz$. Restoring beams used are illustrated in the bottom left corners (same dimensions as in \autoref{fig:g313_7654contour}) and contours are $(-3,3,5,7)\times\sigma$ where $\sigma$ is the noise in the image indicated in the top-left corner of each sub-plot. The green cross indicates a CH$_3$OH maser detected in the $6\GHz$ band whose positional uncertainties are $<0.1\arcsec$.}
\label{fig:g313_7654_F}
\end{center}
\end{figure}

\clearpage
\section{Precession Model}
\label{sec:appJetModel}
\begin{figure}
\begin{center}
\includegraphics[width=.5\textwidth]{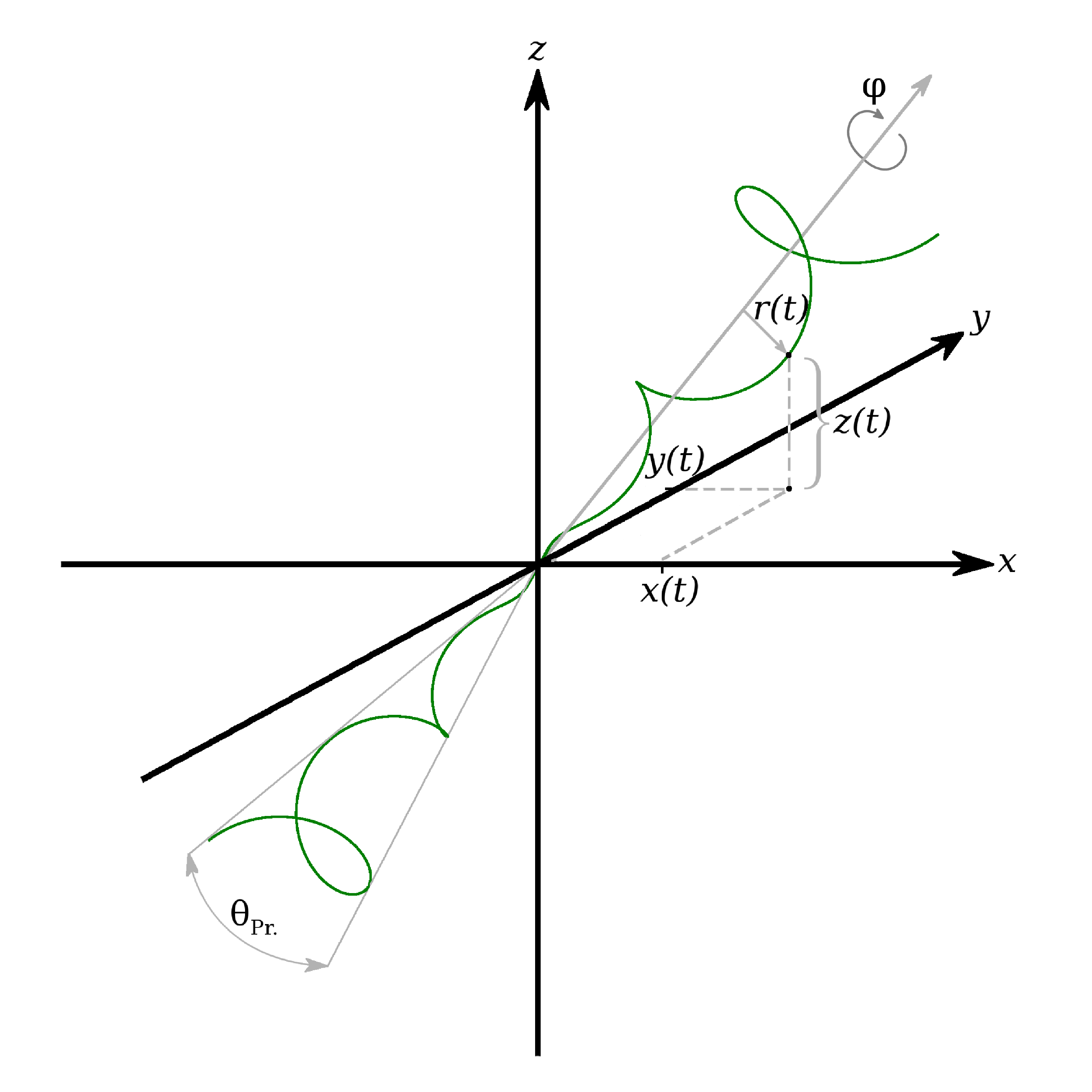} 
\caption{An illustration showing the basic setup of the jet model used to fit radio lobe positional data in the main text. Symbols in the diagram are as follows, $i$ is inclination angle, $\theta_{PA}$ is position angle in the sky, $\theta_{Pr.}$ is the precession angle and $x(t)$, $y(t)$ and $z(t)$ are the $(x,y,z)$ coordinates of a point in the jet's stream (green line) at a time, $t$.}
\label{fig:jetmodel}
\end{center}
\end{figure}

A basic illustration of the model used for a precessing jet is presented in \autoref{fig:jetmodel}, which uses the following assumptions:
\begin{itemize}
\item Ballistic trajectories are adhered to.
\item Bi-axial symmetry of the jet at the point of collimation and ejection.
\item Velocity assumed to be constant along the length of the jet.
\item Precession occurs at a constant angle and period.
\end{itemize}

\begin{align}
    r(t) &=  v\,  t\,  \sin\left(\frac{\theta_{Pr.}}{2}\right)\\
    x(t) &=  r(t)\, \cos\left(\frac{2 \pi t}{P_{Pr.}}\right)\\
    y(t) &=  r(t)\,  \sin\left(\frac{2 \pi t}{P_{Pr.}}\right)\\
    z(t) &=  v\,  t\,  \cos\left(\frac{\theta_{Pr.}}{2}\right)
\end{align}

Once the $(x,y,z)$ coordinates of the jet are established for a position angle of $0\degr$ and inclination of $0\degr$, a rotation is then applied around $x$, $y$ and $z$-axes for position angle, inclination and present jet angle ($\theta_{PA}$, $i$ and $\phi$) respectively. For any point on the jet, $p(x,y,z)$, the final rotated coordinate, $p_{rot}$, is calculated according to \autoref{eq:roteq}:
\begin{align}
R_x(\theta_{PA}) &= \left[ 
\begin{array}{ccc}
1 &      0       &       0     \\
0 & \cos (\theta_{PA})     & -\sin (\theta_{PA})   \\
0 & \sin (\theta_{PA})     &  \cos (\theta_{PA})   
\end{array} \right]\\
R_y(i) &= \left[ 
\begin{array}{ccc}
 \cos(i) & 0 & \sin(i) \\
      0            & 1 &      0            \\
-\sin(i) & 0 & \cos(i) 
\end{array} \right]\\
R_z(\phi) &= \left[ 
\begin{array}{ccc}
\cos(\phi) & -\sin(\phi) & 0 \\
\sin(\phi) &  \cos(\phi) & 0 \\
     0     &       0     & 1 
\end{array} \right]\\
p_{rot} &= \begin{array}{c}
R_y(\theta_{PA}) \cdot(R_x(i) \cdot (R_z(\phi) \cdot p(x,~y,~z))
\label{eq:roteq}
\end{array}
\end{align}

Axes are defined so that the $x$ axis points towards the observer and, therefore, $y$ and $z$ axes represent right ascension and declination respectively. This allows us to compare the observables $r$ and $\theta$ (radial distance and position angle from the jet origin, respectively) with the model, since the $(y,z)$ coordinate of any point in the rotated model is equivalent to $(\alpha,\delta)$ offsets from the jet's origin. Due to the assumption of bi-axial symmetry, it is possible to rotate any lobe coordinate through $\pi$ radians in position angle.

For fitting observational data, a number of models are produced using a range of values for $i$, $\theta_{Pr.}$, $\phi$, $P_{Pr.}$ and $\theta_{PA}$. For each model the reduced $\chi^2$ value is calculated according to the following equation:

\begin{align}
{\chi_\nu}^2 &= \frac{1}{\nu} \sum_{i=0}^{n} \frac{\left(|x_i-\mu_i|-Y\right)^2}{{\sigma_i}^2},
\label{eq:chisquared}
\end{align}

\noindent where $x_i$, $\mu_i$, $\sigma_i$, $\nu$ and $Y$ are the measured value, model value, measurement error,degrees of freedom and Yates' correction factor ($Y=0.5$ if $\nu=1$, otherwise $Y=0$) respectively. Over all tested models, the one which minimizes ${\chi_\nu}^2$ is chosen. From jets with multiple lobes, it is possible to place a lower limit on $\theta_{Pr.}$, approximate $\theta_{PA}$ and in some circumstances place an upper limit on $P_{Pr.}$ (should the lobes appear to move through $>\frac{3\pi}{2}\,\mathrm{rads}$) from inspection of the $(r,\theta)$ lobe coordinates. This allows us to constrain the extent of parameter space covered during our fitting process. 

A shortcoming of this approach is that any number of models can be fitted through any number of points by reducing $P_{Pr.}$ to shorter times. Therefore we choose the simplest model with the longest precession period as the `correct' one. 

\bsp
\label{lastpage}
\end{document}